\documentclass[letterpaper,twocolumn,prl,aps,superscriptaddress,amsmath,amssymb,floatfix]{revtex4}
\usepackage[latin9]{inputenc}
\usepackage{color}
\usepackage{verbatim}
\usepackage{float}
\usepackage{amsmath}
\usepackage{amssymb}
\usepackage{graphicx}
\usepackage{esint}
\usepackage[commandnameprefix=ifneeded]{changes}

\usepackage[unicode=true, bookmarks=true,bookmarksnumbered=false,bookmarksopen=false,
 breaklinks=false,pdfborder={0 0 1},backref=false,colorlinks=true]
 {hyperref}
\hypersetup{ linkcolor=magenta,urlcolor=blue,citecolor=blue,pdfstartview={FitH},hyperfootnotes=false}

\makeatletter

\usepackage{textcomp}
\usepackage{epstopdf}

\pdfpageheight\paperheight
\pdfpagewidth\paperwidth

\@ifundefined{textcolor}{}{%
 \definecolor{BLACK}{gray}{0}
 \definecolor{WHITE}{gray}{1}
 \definecolor{RED}{rgb}{1,0,0}
 \definecolor{GREEN}{rgb}{0,1,0}
 \definecolor{BLUE}{rgb}{0,0,1}
 \definecolor{CYAN}{cmyk}{1,0,0,0}
 \definecolor{MAGENTA}{cmyk}{0,1,0,0}
 \definecolor{YELLOW}{cmyk}{0,0,1,0}
}

\usepackage{xcolor}
\usepackage{soul}
\newcommand{\ket}[1]{\ensuremath{\left|#1\right\rangle}}

\definecolor{blue}{rgb}{0,0,1}
\definecolor{red}{rgb}{1,0,0}
\definecolor{green}{rgb}{0,1,0}

\usepackage{soul}

\makeatother

\begin{document}
\setlength{\belowcaptionskip}{0pt}
\raggedbottom
\title{Impact of structured noise on bosonic quantum error correction}

\author{Guoyu Shu}
\thanks{These authors contributed equally to this work.}
\affiliation{Ministry of Education Key Laboratory for Nonequilibrium Synthesis and Modulation of Condensed Matter, Shaanxi Province Key Laboratory of Quantum Information and Quantum Optoelectronic Devices, School of Physics, Xi'an Jiaotong University, Xi'an 710049, China}
\affiliation{International Quantum Academy, Shenzhen 518048, China}

\author{Ling Hu}
\email{huling@iqasz.cn}
\thanks{These authors contributed equally to this work.}
\affiliation{International Quantum Academy, Shenzhen 518048, China}
\affiliation{Shenzhen Branch, Hefei National Laboratory, Shenzhen 518048, China}

\author{Hongbo Wu}
\affiliation{International Quantum Academy, Shenzhen 518048, China}
\affiliation{Southern University of Science and Technology, Shenzhen 518055, China}

\author{Zhongchu Ni}
\affiliation{International Quantum Academy, Shenzhen 518048, China}

\author{Yanyan Cai}
\affiliation{International Quantum Academy, Shenzhen 518048, China}
\affiliation{Southern University of Science and Technology, Shenzhen 518055, China}

\author{Xiaowei Deng}
\affiliation{International Quantum Academy, Shenzhen 518048, China}

\author{Pan Zheng}
\affiliation{International Quantum Academy, Shenzhen 518048, China}

\author{Pengbo Li}
\email{lipengbo@mail.xjtu.edu.cn}
\affiliation{Ministry of Education Key Laboratory for Nonequilibrium Synthesis and Modulation of Condensed Matter, Shaanxi Province Key Laboratory of Quantum Information and Quantum Optoelectronic Devices, School of Physics, Xi'an Jiaotong University, Xi'an 710049, China}

\author{Fei Yan}
\email{yanfei@baqis.ac.cn}
\affiliation{Beijing Key Laboratory of Fault-Tolerant Quantum Computing, Beijing Academy of Quantum Information Sciences, Beijing 100193, China}

\author{Song Liu}
\affiliation{International Quantum Academy, Shenzhen 518048, China}
\affiliation{Shenzhen Branch, Hefei National Laboratory, Shenzhen 518048, China}

\author{Dapeng Yu}
\affiliation{International Quantum Academy, Shenzhen 518048, China}
\affiliation{Shenzhen Branch, Hefei National Laboratory, Shenzhen 518048, China}

\author{Yuan Xu}
\email{xuyuan@iqasz.cn}
\affiliation{International Quantum Academy, Shenzhen 518048, China}
\affiliation{Shenzhen Branch, Hefei National Laboratory, Shenzhen 518048, China}

\begin{abstract}
Bosonic quantum error correction relies on an accurate description of single-photon loss dissipation using an ideal annihilation operator, under which correctable loss errors act equivalently on different logical code states. In realistic weakly nonlinear bosonic modes, however, coupling to a structured environment can render photon loss Fock-state dependent, thereby modifying the effective dissipation operator and deforming the associated error space. Here, we experimentally investigate this effect in a three-dimensional circuit quantum electrodynamics system. By injecting spectrally engineered environmental noise into a weakly nonlinear oscillator, we observe pronounced frequency selectivity in the measured Fock-state dynamics. While the long-time transition rates converge to the Markovian limit, the transient dynamics display distinctly non-Markovian behavior, in quantitative agreement with a second-order time-convolutionless theory. By applying narrowband noise to a binomially encoded logical qubit, we further demonstrate Fock-state-dependent dissipation channels that drive logical states beyond the nominally correctable error space. These results establish structured environmental noise as a mechanism for deforming photon-loss errors in weakly nonlinear bosonic systems and highlight the need to incorporate realistic, state-dependent dissipation into the design and characterization of bosonic quantum error correction.
\end{abstract}
\maketitle

High-dimensional quantum systems, such as bosonic modes in high-quality superconducting microwave cavities, provide a hardware-efficient platform for quantum information processing~\cite{Braunstein2005,Ma2021,Cai2021,Joshi2021}. A variety of quantum error correction (QEC) codes built on these bosonic modes---including binomial~\cite{Michael2016,Hu2019,Ma2020,Ni2023}, cat~\cite{Mirrahimi2014,Ofek2016}, and Gottesman-Kitaev-Preskill codes~\cite{Gottesman2001,Fluhmann2019,CampagneIbarcq2020,Sivak2023}---have been explored, and recent experiments have demonstrated bosonic QEC beyond the break-even point~\cite{Ofek2016,Ni2023,Sivak2023,Sun2025,Ni2025,Shirol2026} as well as concatenated encoding architectures~\cite{Putterman2025,Noh2020,Noh2022}. Bosonic QEC typically assumes a photon-loss model described by an ideal annihilation operator $\hat{a}$ of a harmonic oscillator mode, indicating that the loss rates follow the expected bosonic scaling with photon number~\cite{Ma2021}. Under this assumption, correctable loss errors act equivalently on different logical code states, allowing them to be identified and corrected within a common error-syndrome structure~\cite{Michael2016,Albert2018}.

\begin{figure}[tbhp]
\centering
\includegraphics{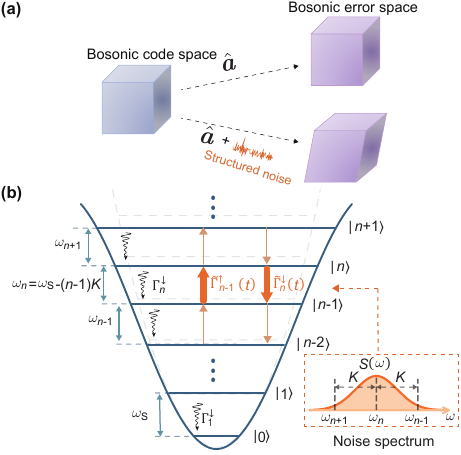} 
\caption{\textbf{Bosonic QEC under structured noise and Fock-state-dependent dissipation.}
(a) Schematic illustration of the effect of structured noise on bosonic QEC. Conventional photon loss maps the logical state from the code space to a correctable error space, whereas structured noise modifies the photon-loss error structure, thereby deforming the error space.
(b) Energy level diagram of the weakly nonlinear bosonic mode with frequency $\omega_\mathrm{S}$ and Kerr nonlinearity $K$. Adjacent Fock-state transitions occur at $\omega_n=\omega_\mathrm{S}-(n-1)K$, with intrinsic decay rate $\Gamma_n^{\downarrow}$. A narrowband noise spectrum $S(\omega)$ centered at $\omega_n$ selectively induces excitation and relaxation between spectrally covered Fock states (orange arrows), with the corresponding transition rates satisfying $\widetilde{\Gamma}_{n-1}^{\uparrow}(t)=\widetilde{\Gamma}_{n}^{\downarrow}(t)$.
}
\label{fig:fig1}
\end{figure}

However, realistic bosonic systems inevitably exhibit weak nonlinearities arising from coupling to qubits, nonlinear elements, or control components ~\cite{Blais2004,Blais2021,Chapman2023}, which lift the degeneracy of adjacent Fock-state transitions~\cite{Kirchmair2013,Grimm2020}. In structured environments associated with quasiparticle excitations, two-level fluctuators, or localized defects in superconducting quantum devices~\cite{Martinis2005,Simmonds2004,Catelani2011}, the frequency dependence of environmental noise can lead to Fock-state-dependent dissipation~\cite{Clerk2010}. This modifies the photon-loss error structure assumed in bosonic QEC and deforms the corresponding error space [Fig.~\ref{fig:fig1}(a)]. Understanding this interplay between bosonic nonlinearities and environmental spectral structures is therefore essential for accurately modeling and optimizing bosonic quantum information processors. However, microscopic noise sources in solid-state quantum devices are difficult to isolate and quantitatively characterize~\cite{Martinis2005,Martinis2009,Siddiqi2021,Yan2018}, and the impact of structured environmental spectra on Fock-state-dependent dissipation remains unexplored. Engineered noise with tunable spectral properties provides a controlled way to investigate these effects.

In this work, we experimentally investigate Fock-state-dependent dissipation induced by structured noise in a weakly nonlinear bosonic system, and develop a theoretical framework to capture the resulting non-Markovian dynamics. We derive a second-order time-convolutionless (TCL2) master equation~\cite{Breuer2002} that describes the time-dependent transition rates between adjacent Fock states and establishes their connection to the noise spectral density and finite correlation time. We experimentally validate these predictions in a three-dimensional (3D) superconducting microwave cavity by applying engineered narrowband noise with tunable frequency and power. The measured dynamics exhibit both frequency selectivity and transient non-Markovian behavior, while the long-time-limit transition rates recover the expected bosonic scaling with photon number, approaching the Markovian regime. Finally, using a binomial logical qubit, we show that narrowband structured noise gives rise to Fock-state-dependent error channels that deviate from the conventional photon-loss model assumed in ideal bosonic QEC, thereby deforming the correctable error space.
Our results provide a framework for investigating how structured environments affect bosonic dissipation dynamics and for assessing their implications for bosonic QEC.

We first consider a quantum harmonic oscillator with a weak Kerr-nonlinearity, whose energy level diagram is illustrated in Fig.~\ref{fig:fig1}(b). Setting $\hbar=1$, the bosonic mode Hamiltonian is
\begin{equation}
H_{\mathrm{S}} = \omega_\mathrm{S} \hat{n} - \frac{K}{2} \hat{n}(\hat{n}-1),
\end{equation}
where $\omega_\mathrm{S}$ is the mode frequency, $\hat{n}=\hat{a}^\dagger\hat{a}$ is the photon number operator, and $K$ is the self-Kerr nonlinearity. The Kerr nonlinearity induces anharmonicity in the spectrum, so that the transition between adjacent Fock states $|n\rangle$ and $|n-1\rangle$ has a different resonance frequency $\omega_n=\omega_\mathrm{S}-K(n-1)$. We denote the intrinsic decay rate of this transition by $\Gamma_n^{\downarrow}$, with $\Gamma_1^{\downarrow}=1/T_{1, \mathrm{S}}$, where $T_{1, \mathrm{S}}$ is the single-photon lifetime of the bosonic mode.

The bosonic mode is subjected to classical structured noise $\xi(t)$ [orange color in Fig.~\ref{fig:fig1}(b)], characterized by a power spectral density $ S(\omega)=\int_{-\infty}^{\infty}d\tau \langle\xi^*(\tau)\xi(0)\rangle e^{i\omega\tau}$~\cite{Clerk2010}, with a correlation time $\tau_{\mathrm{c}}$.
When the interaction time satisfies $t\lesssim\tau_{\mathrm{c}}$, memory effects become non-negligible, giving rise to non-Markovian dynamics. In particular, when $\tau_{\mathrm{c}}\sim T_{1, \mathrm{S}}$, these memory effects persist on the timescale of bosonic decay and become experimentally observable.
To describe the non-Markovian dynamics~\cite{Breuer2016,Brillant2025,Paladino2014,Figueroa2021}, we employ the TCL2 master equation~\cite{Breuer2002} for the reduced density matrix $\tilde{\rho}_{\mathrm{S}}(t)$ of the bosonic mode in the interaction picture:
\begin{equation}
\dot{\tilde{\rho}}_\mathrm{S}(t)
=
-\int_0^t d\tau\,
\left\langle
\left[
\tilde{H}_{\mathrm{I}}(t),
\left[
\tilde{H}_{\mathrm{I}}(t-\tau),
\tilde{\rho}_\mathrm{S}(t)
\right]
\right]
\right\rangle_{\xi}.
\end{equation}
where $H_{\mathrm{I}}(t)=\xi(t)\hat{a}^\dagger+\xi^*(t)\hat{a}$ is the interaction term between the noise and the bosonic mode.
The TCL2 framework retains the finite upper limit $t$ in the temporal integration, resulting in time-dependent transition rates. 

We then reduce the TCL2 dynamics to the population dynamics in the Fock-state basis, which decouple from the off-diagonal coherence terms and can be described by a Lindblad-like rate equation:
\begin{equation}
\begin{aligned}
\left.\dot{\tilde{\rho}}_{\mathrm{S}}(t)\right|_{\mathrm{diag}}
=& \sum_{n=1}^\infty \left[
\tilde{\gamma}_{n}^{\downarrow}(t)\mathcal{D}[\hat{A}_n]
+ \tilde{\gamma}_{n-1}^{\uparrow}(t)\mathcal{D}[\hat{A}_n^\dagger]
\right]\tilde{\rho}_{\mathrm{S}}(t)\\
&+ \Gamma_{1}^{\downarrow}\mathcal{D}[\hat{a}]\tilde{\rho}_{\mathrm{S}}(t).
\end{aligned}
\end{equation}
Here, $\hat{A}_n=\sqrt{n}|n-1\rangle\langle n|$ is the transition operator between adjacent Fock states, and $\mathcal{D}[\hat{\mathcal{O}}]\tilde{\rho}_{S}=\hat{\mathcal{O}}\tilde{\rho}_{S}\hat{\mathcal{O}}^\dagger-\frac{1}{2}\{\hat{\mathcal{O}}^\dagger\hat{\mathcal{O}},\tilde{\rho}_{S}\}$. $\tilde{\gamma}_{n}^{\downarrow}(t)$ and $\tilde{\gamma}_{n-1}^{\uparrow}(t)$ describe the downward and upward transition channels induced by noise, respectively. The last term accounts for the intrinsic decay.

The effective time-dependent downward transition rate from $\ket{n}$ to $\ket{n-1}$ incorporates the bosonic enhancement factor $n$ and takes the form:
\begin{equation}
\tilde{\Gamma}_{n}^{\downarrow}(t)=n\tilde{\gamma}_{n}^{\downarrow}(t)=n\int^{+\infty}_{-\infty}d\omega' S(\omega')\frac{\sin[(\omega'-\omega_n)t]}{\pi(\omega'-\omega_n)}.
\label{eq:NM_rate}
\end{equation}
For a symmetric spectrum $S(\omega)=S(-\omega)$, the upward and downward rates coincide, $\tilde{\Gamma}_{n-1}^{\uparrow}(t)=\tilde{\Gamma}_{n}^{\downarrow}(t)$, and the sinc function acts as a finite-time filter. When $t\lesssim\tau_\mathrm{c}$, the broad filter admits off-resonant noise components, giving rise to transient oscillations and memory effects.
In the long-time regime ($t\gg\tau_{\mathrm{c}}$), the sinc filter approaches $\delta(\omega'-\omega_n)$, and the transition rate converges to the Markovian value determined by $S(\omega_n)$~\cite{Clerk2010} (see Supplemental Sec.~II-A):
\begin{equation}
\tilde{\Gamma}_{n}^{\downarrow,\mathrm{L}}=nS(\omega_n).
\label{eq:M_rate}
\end{equation}

\begin{figure}[tbhp]
\centering
\includegraphics{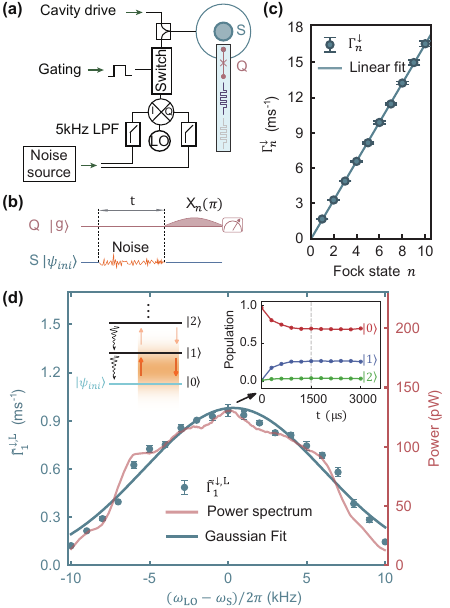} \caption{\textbf{Experimental setup and characterization of structured noise.}
(a) Noise-generation setup. Two independent noise signals are low-pass filtered, IQ up-converted with a tunable local oscillator (LO), gated by a microwave switch, and combined with the cavity drive.
(b) Experimental sequence. The initial state $\ket{\psi_{\mathrm{ini}}}$ is prepared in the bosonic mode ($\mathrm{S}$), followed by a noise pulse of duration $t$ and photon-number-selective measurement with the ancilla qubit ($\mathrm{Q}$).
(c) Intrinsic Fock-state decay rates $\Gamma_n^\downarrow$ (dots). The linear fit (line) follows the expected bosonic scaling $\Gamma_n^\downarrow=n\Gamma_1^\downarrow$.
(d) Effective noise spectrum. The red line shows the noise spectrum measured with a spectrum analyzer. The blue dots show $\widetilde{\Gamma}_{1}^{\downarrow,\mathrm{L}}$ extracted from steady-state populations with $\ket{\psi_{\mathrm{ini}}}=\ket{0}$ while sweeping $\omega_{\mathrm{LO}}$, consistent with the measured noise spectrum; the blue line is a Gaussian fit. Insets show the noise-induced transitions (left) and representative population dynamics at $\omega_{\mathrm{LO}}=\omega_\mathrm{S}$ (right), with the steady-state populations obtained by averaging over $t>1.5~\mathrm{ms}$. Error bars denote the standard deviation of repeated experiments.
}
\label{fig:fig2}
\end{figure}

In our experiment, the bosonic mode is constructed by a 3D superconducting microwave cavity, which is dispersively coupled to a superconducting transmon qubit~\cite{Blais2021,Joshi2021,Cai2021,Reagor2016,Axline2016}. 
The dispersive interaction, with a coupling strength of $\chi/2\pi=2.65~\mathrm{MHz}$, enables photon-number-selective measurement and induces a cavity Kerr nonlinearity of $K/2\pi=10.1~\mathrm{kHz}$. The bosonic mode has a single-photon lifetime of $T_{1,\mathrm{S}}=0.6~\mathrm{ms}$ and a Ramsey dephasing time of $T_{2,\mathrm{S}}=1.1~\mathrm{ms}$ (see Supplementary Sec.~I).

With this device, we experimentally investigate the non-Markovian behavior by injecting spectrally engineered environmental noise into the cavity. This tunable narrowband noise is generated using the scheme shown in Fig.~\ref{fig:fig2}(a). Two independent noise signals from an arbitrary waveform generator (AWG) are low-pass filtered at $5~\mathrm{kHz}$ and up-converted through an IQ mixer with a local oscillator (LO), which sets the noise center frequency $\omega_\mathrm{LO}$. The resulting noise is then gated by a microwave switch and combined with the cavity drive~\cite{Wang2026}.
The noise-induced dynamics are characterized using the pulse sequence shown in Fig.~\ref{fig:fig2}(b). The bosonic mode ($\mathrm{S}$) is first prepared in the initial state $\ket{\psi_{\mathrm{ini}}}$, followed by the application of a noise pulse with variable duration $t$. The populations $P_n(t)$ of selected Fock states $\ket{n}$ are subsequently measured using the corresponding photon-number-selective $\pi$ pulses $X_n(\pi)$ on the ancilla qubit. Without noise, the measured intrinsic decay rates of the bosonic mode exhibit the expected bosonic scaling, $\Gamma_n^{\downarrow}= n\Gamma_1^{\downarrow}$~\cite{Wang2008,Peterer2015, Wang2025,Kim2025}, as shown in Fig.~\ref{fig:fig2}(c) (see Supplementary Sec.~IV-D).
 
We first calibrate the effective noise spectrum. Starting from $\ket{\psi_{\mathrm{ini}}}=\ket{0}$, we sweep the noise center frequency $\omega_\mathrm{LO}$ across $\omega_{\mathrm{S}}\pm10~\mathrm{kHz}$, while measuring the Fock-state populations $P_{0,1,2}(t)$ over time. The upper right panel of Fig.~\ref{fig:fig2}(c) shows the population dynamics at $\omega_\mathrm{LO}=\omega_{\mathrm{S}}$. When $t\gg\tau_{\mathrm{c}}$, the noise-induced transition rates become time independent, and with further evolution, the populations of $\ket{0}$ and $\ket{1}$ reach a steady state at $t>1.5~\mathrm{ms}$. From the steady-state populations $P_{0,1}^{\mathrm{ss}}$, we extract the long-time-limit transition rate $\widetilde{\Gamma}_{1}^{\downarrow,\mathrm{L}}=\Gamma_{1}^{\downarrow}\frac{P_1^{\mathrm{ss}}}{P_0^{\mathrm{ss}}-P_1^{\mathrm{ss}}}$. 
Repeating this procedure for each $\omega_\mathrm{LO}$ yields the effective noise spectrum, which agrees well with the noise spectrum measured using a spectrum analyzer [Fig.~\ref{fig:fig2}(c)]. Gaussian fits to independent measurements yield $\sigma/2\pi=5.7\pm0.1~\mathrm{kHz}$ ($\tau_{\mathrm{c}}\sim40~\mu\mathrm{s}$) and a peak rate of $0.98\pm0.01~\mathrm{ms}^{-1}$, where the uncertainties are standard deviations, demonstrating the frequency selectivity of the noise-induced transitions.

\begin{figure*}[tbhp]
\centering
\includegraphics{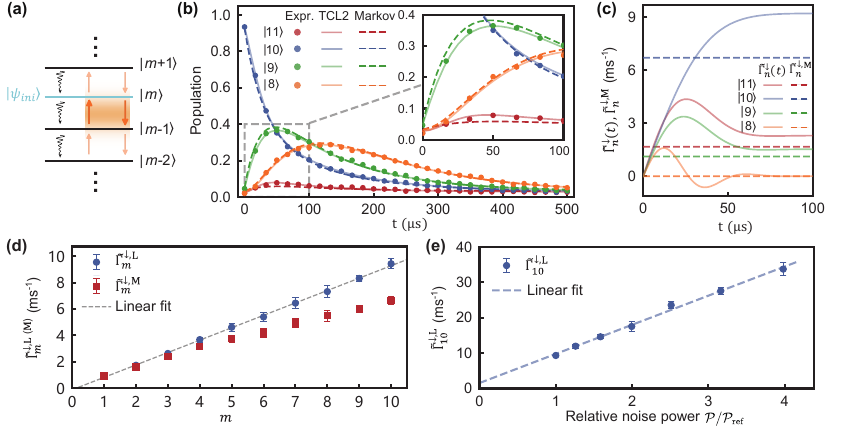} 
\caption{\textbf{Non-Markovian Fock-state dynamics induced by structured narrowband noise.}
(a) Noise-induced Fock-state dynamics for $\ket{\psi_{\mathrm{ini}}}=\ket{m}$ and $\omega_\mathrm{LO}=\omega_m$, with populations $P_n(t)$ measured for $\ket{n}=\ket{m+1},\ldots,\ket{m-2}$.
(b) Population dynamics for $m=10$. Experimental data (dots) are fitted by the TCL2 theory (solid curves) and Markovian model (dashed curves). Inset: zoom-in of the early-time dynamics. 
(c) Transition rates corresponding to the inset of (b): time-dependent TCL2 rates $\widetilde{\Gamma}_n^\downarrow(t)$ (solid) and constant Markovian rates $\widetilde{\Gamma}_n^{\downarrow,\mathrm{M}}$ (dashed).
(d) TCL2 long-time-limit transition rate $\widetilde{\Gamma}_{m}^{\downarrow,\mathrm{L}}$ (blue) and Markovian transition rate $\widetilde{\Gamma}_{m}^{\downarrow,\mathrm{M}}$ (red) versus the initial Fock state $\ket{\psi_{\mathrm{ini}}}=\ket{m}$. Gray dashed line: linear fit to $\widetilde{\Gamma}_{m}^{\downarrow,\mathrm{L}}$.
(e) Repeating the $m=10$ experiment in (b), with the injected noise strength varied by changing the attenuator, showing $\widetilde{\Gamma}_{10}^{\downarrow,\mathrm{L}}$ versus relative noise power $\mathcal{P}/\mathcal{P}_{\mathrm{ref}}$, where $\mathcal{P}_{\mathrm{ref}}$ is the noise power used above. Fit with $f=Ax+B$ (dashed line), yields $A=8.2\pm 0.3~\mathrm{ms}^{-1}$ and $B=1.5\pm0.7 ~\mathrm{ms}^{-1}$. Error bars denote the standard deviation of repeated experiments.}
\label{fig:fig3}
\end{figure*}

We next investigate the non-Markovian Fock-state dynamics induced by the narrowband noise. As illustrated in Fig.~\ref{fig:fig3}(a), the cavity is initialized in a Fock state $\ket{\psi_{\mathrm{ini}}}=\ket{m}$, with the narrowband noise centered at the corresponding transition frequency $\omega_\mathrm{LO}=\omega_m$, and the populations $P_n(t)$ of the Fock states $\ket{n}=\ket{m+1},\ldots,\ket{m-2}$ are measured. We first consider $m=10$ as a representative example. The measured populations $P_n(t)$ are shown in Fig.~\ref{fig:fig3}(b) and are fitted separately using the TCL2 and Markovian master equations. In both fits, the Gaussian noise spectrum is centered at $\omega_\mathrm{LO}$ with $\sigma/2\pi=5.7~\mathrm{kHz}$, and the intrinsic decay rates $\Gamma_n^{\downarrow}$ are fixed to their independently measured values, while the noise strength $\eta$ is the only free parameter. The TCL2 and Markovian fits use Eqs.~\ref{eq:NM_rate} and \ref{eq:M_rate}, respectively, to obtain $P_n(t,\eta)$ and then fit $\mathrm{A}P_n(t,\eta)+\mathrm{C}$ to the experimental populations. Note that the measured noise spectrum has a $0.3~\mathrm{kHz}$ center-frequency offset from $\omega_\mathrm{LO}$, which is included in the fitting (see Supplementary Sec.~III).
As shown in the inset of Fig.~\ref{fig:fig3}(b), the TCL2 theory accurately captures the measured dynamics, particularly the short-time transient behavior, whereas the Markovian model shows clear deviations.
The inferred TCL2 rates $\widetilde{\Gamma}_{n}^{\downarrow}(t)$ exhibit transient oscillations and approach their long-time limits $\widetilde{\Gamma}_{n}^{\downarrow,\mathrm{L}}$ [Fig.~\ref{fig:fig3}(c)]. Since the noise is centered at $\omega_{10}$, $\widetilde{\Gamma}_{10}^{\downarrow,\mathrm{L}}$ is maximal. In contrast, the Markovian fit yields a constant rate $\widetilde{\Gamma}_{n}^{\downarrow,\mathrm{M}}$ averaged over the entire evolution, which thus deviates from $\widetilde{\Gamma}_{n}^{\downarrow,\mathrm{L}}$ (see Supplementary Sec.~II-B).

To examine the photon-number dependence of the noise-induced transition rates, we compare $\widetilde{\Gamma}_{m}^{\downarrow,\mathrm{L}}$ and $\widetilde{\Gamma}_{m}^{\downarrow,\mathrm{M}}$ for $m=1,\ldots,10$, as shown in Fig.~\ref{fig:fig3}(d). $\widetilde{\Gamma}_{m}^{\downarrow,\mathrm{L}}$ scale approximately linearly with $m$, consistent with the bosonic enhancement $|\langle m-1|\hat{a}|m\rangle|^2=m$, whereas $\widetilde{\Gamma}_{m}^{\downarrow,\mathrm{M}}$ deviate from this linear scaling, with the deviation increasing at higher photon numbers as the shortened Fock-state lifetime increases the relative contribution of the non-Markovian transient to the overall decay dynamics (see Supplementary Sec.~II-B). A linear fit to $\widetilde{\Gamma}_{m}^{\downarrow,\mathrm{L}}$ yields a slope of $0.94\pm0.01 ~\mathrm{ms}^{-1}$ and a intercept of $-0.12\pm0.06 ~\mathrm{ms}^{-1}$, consistent with the expected linear scaling with $m$ and the result in Fig.~\ref{fig:fig2}(c). Furthermore, for $m=10$, we vary the injected noise strength at fixed $\omega_\mathrm{LO}=\omega_{10}$ and find that $\widetilde{\Gamma}_{10}^{\downarrow,\mathrm{L}}$ scales approximately linearly with the noise strength [Fig.~\ref{fig:fig3}(e)], consistent with Eq.~\ref{eq:M_rate}. Overall, the TCL2 framework captures the non-Markovian dynamics and their long-time Markovian limit.

Finally, we examine the impact of narrowband noise on a bosonic logical qubit using the lowest-order binomial code, with $\ket{0_\mathrm{L}}=(\ket{0}+\ket{4})/\sqrt{2}$ and $\ket{1_\mathrm{L}}=\ket{2}$. Without applied noise, single-photon loss maps the logical states to the correctable error space spanned by $\ket{1}$ and $\ket{3}$. After preparing $\ket{\psi_{\mathrm{ini}}}=\ket{+}_{\mathrm{L}}=(\ket{0}_{\mathrm{L}}+\ket{1}_{\mathrm{L}})/\sqrt{2}$, we set $\omega_\mathrm{LO}=\omega_4$ and monitor the populations of $\ket{1}$ and $\ket{3}$ to probe the error-space dynamics. As shown in Fig.~\ref{fig:fig4}(a), without noise, their populations follow the correctable dynamics under intrinsic photon loss and no-jump evolution, whereas narrowband noise strongly modifies the $\ket{3}$ component while leaving the $\ket{1}$ component nearly unchanged, revealing distinct dissipation dynamics of the two logical components. Such noise-induced Fock-state-dependent dissipation deforms the correctable error space, violates the QEC condition, and ultimately leads to uncorrectable loss of logical information. Detuning the noise away from $\omega_4$ restores the populations to their free-evolution values [Fig.~\ref{fig:fig4}(b)], demonstrating the frequency selectivity of this effect. These results show that Fock-state-dependent dissipation can violate the QEC condition and degrade bosonic QEC performance.

In conclusion, we have experimentally and theoretically investigated Fock-state-dependent dissipation induced by structured environmental noise in a weakly nonlinear bosonic system, and examined its implications for bosonic QEC. By injecting spectrally engineered narrowband noise into a weakly nonlinear Kerr cavity and resolving individual Fock-state transitions, we observe transient non-Markovian dynamics that are quantitatively described by the TCL2 theory. Within this framework, the effective transition rates between adjacent Fock states become time-dependent and explicitly depend on the noise spectral density and its finite correlation time. Using a binomially encoded logical qubit, we further showed that structured narrowband noise induces Fock-state-dependent dissipation, modifies the error dynamics of the logical states, and deforms the correctable error space. 

These results establish Fock-state-dependent dissipation channels induced by structured environmental noise as a mechanism for deforming photon-loss errors in bosonic QEC~\cite{Ma2021,Michael2016,Albert2018}. Our work provides a controlled platform to investigate spectrally engineered dissipation~\cite{Harrington_NRP2022_4_660} in structured environments and its non-Markovian effects in bosonic systems~\cite{Udupa2026}. Understanding how environmental spectral structure is mapped onto logical error channels may guide the design of bosonic encoding and error-correction strategies adapted to realistic noise environments. Looking forward, tailoring the noise spectrum could selectively suppress the decay of logical states relative to error states, potentially realizing error-transparent or error-biased codes that could relax error-correction requirements~\cite{Webster2015PRA_92_062309}. In addition, extending our approach to multi-tone noise injection would enable simultaneous control of multiple dissipation channels, allowing the synthesis of complex non-equilibrium steady states and the exploration of quantum thermodynamic phenomena~\cite{Deffner_book_2019}. 

\begin{figure}[t]
\centering
\includegraphics{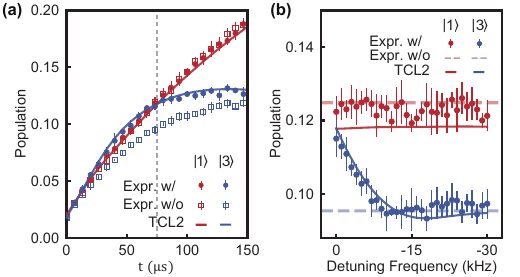} 
\caption{\textbf{Effect of structured narrowband noise on a bosonic logical qubit.} (a) Population dynamics of $|1\rangle$ and $|3\rangle$ with and without applied noise after preparing the $|+\rangle_\mathrm{L}$ state of the lowest-order binomial code. Filled circles and open squares denote evolution with narrowband noise centered at $\omega_4$ and without noise, respectively. (b) Populations of $|1\rangle$ and $|3\rangle$ at $t=75~\mu\mathrm{s}$ as a function of the noise center frequency $\omega_\mathrm{LO}$. As the noise is detuned from $\omega_4$, the populations recover toward their corresponding values without noise (dashed lines). The TCL2 curves are calculated using the fitted parameters from Fig.~\ref{fig:fig3}(d) and the experimental initial state. Error bars denote the standard deviation of repeated experiments.}
\label{fig:fig4}
\end{figure}

\textit{Acknowledgments}---This work was supported by the National Natural Science Foundation of China (Grants No.~12422416, No.~12274198, No.~12374471,
No.~12575025,
No.~12604759,
No.~92476206 and No.~12322413), the Innovation Program for Quantum Science and Technology (Grants No.~2024ZD0302300, No.~2021ZD0301703), the Guangdong Basic and Applied Basic Research Foundation (Grant No.~2024B1515020013), the Shenzhen Science and Technology Program (Grants No.~RCYX20221008092907026 and No. QNXMA20250701092312018), and the Beijing Natural Science Foundation (Grants No. JQ25014).


%

\clearpage{}

\newpage   
\clearpage 
\onecolumngrid

\clearpage

\setcounter{section}{0}

 \begin{center}
  \large\bfseries
  Supplementary Material for ``Impact of Structured Noise on Bosonic Quantum Error Correction''
\end{center}

\maketitle


\section{Experimental Device and System Parameters}
\subsection{Experimental device and setup}

\begin{figure*}[htbp]
    \centering
    \includegraphics[width=\linewidth]{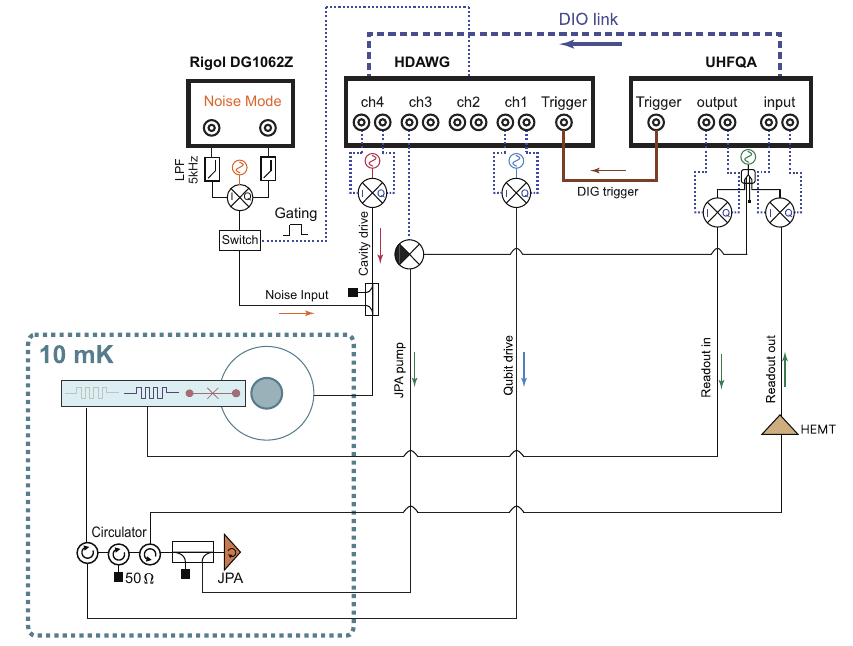}
    \caption{\textbf{Experimental setup.} }
    \label{fig:SM_Setup}
\end{figure*}

The experiment is performed in a three-dimensional (3D) circuit quantum electrodynamics (cQED) architecture. The device consists of a 3D coaxial cavity machined from high-purity aluminum~\cite{Axline2016}, serving as the principal bosonic mode. Coupled to this storage cavity is an auxiliary transmon qubit fabricated on a sapphire substrate. The transmon features an $\text{Al}-\text{Al}_2\text{O}_3-\text{Al}$ Josephson junction, with antenna pads and readout striplines patterned from body-centered cubic (BCC) $\alpha$-phase tantalum films~\cite{Place2021}. A Purcell-filtered stripline resonator is strongly coupled to the transmon for dispersive state readout.

The device is thermalized below 10 mK inside a dilution refrigerator and shielded against environmental magnetic fields. The bosonic mode and auxiliary qubit are manipulated via IQ-modulated microwave pulses. These control pulses are synthesized using high-density arbitrary waveform generators (ZI HDAWG), and the readout is orchestrated by an ultra-high frequency quantum analyzer (ZI UHFQA).

To investigate the Fock-state dynamics induced by structured noise, we generate a narrowband Gaussian noise field using two independent stochastic waveforms produced by an arbitrary waveform generator (RIGOL DG1062Z). Two independent signals form the I and Q components of a complex Gaussian noise field, with their dc offsets adjusted to suppress LO leakage. The two baseband signals are passed through 5~kHz low-pass filters, which define the noise bandwidth and the corresponding correlation timescale, and are subsequently upconverted by an IQ mixer driven by a microwave LO. The LO frequency sets the center frequency $\omega_{\mathrm{LO}}$ of the noise spectrum, allowing the structured noise to be tuned across different Kerr-resolved transitions of the bosonic mode without changing its baseband spectral profile. A microwave switch, gated by a marker signal from the HDAWG, defines the noise-injection window. The noise drive is then combined with the coherent cavity-control pulses through a directional coupler before being delivered to the storage cavity.

\subsection{System Hamiltonian}

Our experimental system consists of a 3D superconducting microwave cavity coupled to a transmon ancilla and a readout resonator. The cavity serves as the bosonic mode, while the ancilla enables state preparation and measurement. The weak Kerr nonlinearity of the bosonic mode spectrally resolves adjacent Fock-state transitions, allowing us to selectively probe Fock-state dynamics under structured noise. In the dispersive regime, the system is described by the effective Hamiltonian~\cite{Blais2021,Cai2021}:
\begin{equation}
\begin{aligned}
\hat{H}/\hbar &= \omega_q \hat{q}^\dagger \hat{q} + \omega_S \hat{a}^\dagger \hat{a} + \omega_r \hat{r}^\dagger \hat{r} \\
&\quad - \frac{K_q}{2} \hat{q}^{\dagger 2} \hat{q}^2 - \frac{K}{2} \hat{a}^{\dagger 2} \hat{a}^2 \\
&\quad - \chi_{qr} \hat{q}^\dagger \hat{q} \hat{r}^\dagger \hat{r} - \chi_{qS} \hat{q}^\dagger \hat{q} \hat{a}^\dagger \hat{a}.
\end{aligned}
\label{eq:hamiltonian}
\end{equation}
Here, $\omega_S$ ($\omega_q, \omega_r$) and $\hat{a}$ ($\hat{q}, \hat{r}$) denote the frequency and annihilation operator of the storage mode (transmon ancilla, readout resonator). The parameters $K_q$ and $K$ are the respective self-Kerr nonlinearities, while $\chi_{qS}$ and $\chi_{qr}$ quantify the dispersive shifts of the ancilla to the storage cavity and readout resonator.

The key device parameters and coherence properties measured in this experiment are summarized in Table~\ref{tab:params}.

\begin{table}[h]
    \centering
    \caption{Device parameters and coherence properties.}
    \label{tab:params}
    \begin{tabular}{lc}
        \hline \hline
        \textbf{Parameters} & \textbf{Values} \\
        \hline
        Cavity frequency $\omega_S/2\pi$ (GHz) & 6.532 \\
        Qubit frequency $\omega_q/2\pi$ (GHz) & 4.962 \\
        Readout frequency $\omega_r/2\pi$ (GHz) & 8.562 \\
        $Q$-$S$ dispersive shift $\chi_{qS}/2\pi$ (MHz) & 2.65 \\
        $Q$-$R$ dispersive shift $\chi_{qr}/2\pi$ (MHz) & 1.9 \\
        Qubit anharmonicity $K_q/2\pi$ (MHz) & 216 \\
        Cavity anharmonicity $K/2\pi$ (kHz) & 10.1 \\
        Qubit $T_{1q}$ ($\mu$s) & 98 \\
        Qubit $T_{2q}$ ($\mu$s) & 120 \\
        Cavity $T_{1S}$ ($\mu$s) & 602 \\
        Cavity $T_{2S}$ ($\mu$s) & 1080 \\
        Readout resonator $T_{1r}$ (ns)  & 62 \\
        \hline \hline
    \end{tabular}
\end{table}

\section{Theory}
\subsection{TCL2 Master Equation}

In this section, we derive the second-order time-convolutionless (TCL2) master equation for a weakly nonlinear bosonic mode subjected to structured noise. The total Hamiltonian is given by
\begin{equation}
H = H_S + H_B + H_{\mathrm{int}},
\end{equation}
where $H_S$, $H_B$, and $H_{\mathrm{int}}$ denote the system, bath, and system--bath interaction Hamiltonians, respectively.

The system Hamiltonian $H_S$ is given by~\cite{Kirchmair2013}
\begin{equation}
H_S/\hbar = \omega_S a^\dagger a - \frac{K}{2}a^{\dagger 2}a^2
= \omega_S \hat{n} - \frac{K}{2}\hat{n}(\hat{n}-1),
\label{Hamiltonian}
\end{equation}
where $\hat{n}=a^\dagger a$ is the photon-number operator, $\omega_S$ is the resonance frequency of the bosonic mode, and $K$ denotes the Kerr nonlinearity. The transition frequency between adjacent Fock states $\vert n\rangle$ and $\vert n-1\rangle$ is therefore
\begin{equation}
\omega_n = \omega_S - K(n-1).
\end{equation}

The bath and interaction Hamiltonians take the standard bilinear form~\cite{Clerk2010}
\begin{equation}
H_B = \int d\omega\, \omega b^\dagger(\omega)b(\omega),
\end{equation}
and
\begin{equation}
H_{\mathrm{int}} = a^\dagger B + a B^\dagger,
\label{eq:int}
\end{equation}
where $B = \int d\omega\, g(\omega)b(\omega)$ is the effective bath operator, where $g(\omega)$ is the frequency-dependent coupling strength.

In the interaction picture, the annihilation operator of the weakly nonlinear bosonic mode can be decomposed into individual transitions along the Fock-state ladder:
\begin{equation}
\tilde{a}(t)
= e^{iH_St} a e^{-iH_St}
= \sum_{n=1}^{\infty} A_n e^{-i\omega_n t},
\label{eq:a_A}
\end{equation}
where $A_n = \sqrt{n}\vert n-1\rangle\langle n\vert$
is the transition operator associated with the $\vert n\rangle\leftrightarrow\vert n-1\rangle$ transition, and the corresponding transition frequency $\omega_n$ is shifted by the Kerr nonlinearity. The bath operator evolves as $\tilde{B}(t)=e^{iH_Bt}Be^{-iH_Bt}$.

To describe the dynamics induced by structured noise with a finite correlation time $\tau_c$, we employ the second-order time-convolutionless (TCL2) master equation without invoking the Markov approximation. The distinction between the finite-time TCL2 description and the Markovian limit is governed by the relation between the noise interaction time $t$ and the noise correlation time $\tau_c$. When $t$ is comparable to $\tau_c$~\cite{Clerk2010,Breuer2016,Brillant2025,Breuer2002}, the system retains information about the finite temporal correlations of the noise, giving rise to transient non-Markovian dynamics. In contrast, for $t\gg\tau_c$, the noise correlations decay on a timescale much shorter than the evolution time, and the Markov approximation becomes applicable.

Within the TCL2 approximation, the interaction-picture equation of motion is~\cite{Breuer2002}:
\begin{equation}
\frac{d\tilde{\rho}_S(t)}{dt}
=
-\int_0^t d\tau\,
\operatorname{Tr}_B
\left[
  \tilde{H}_{\mathrm{int}}(t),
  \left[
    \tilde{H}_{\mathrm{int}}(t-\tau),
    \tilde{\rho}_S(t)\otimes\rho_B
  \right]
\right].
\end{equation}
Here, $\rho_B$ denotes the vacuum state of the quantum bath responsible for intrinsic single-photon loss. The classical stochastic drive is treated separately by averaging over noise realizations, as in Eq.~(2) of the main text.

Expanding the double commutator and tracing over the bath degrees of freedom gives
\begin{equation}
\begin{aligned}
\frac{d\tilde{\rho}_S}{dt}
&=
\int_0^t d\tau\,
\langle B(\tau)B^\dagger(0)\rangle
\left[
a(t-\tau)\tilde{\rho}_S a^\dagger(t)
-
a^\dagger(t)a(t-\tau)\tilde{\rho}_S
\right]
\\
&\quad+
\int_0^t d\tau\,
\langle B^\dagger(\tau)B(0)\rangle
\left[
a^\dagger(t-\tau)\tilde{\rho}_S a(t)
-
a(t)a^\dagger(t-\tau)\tilde{\rho}_S
\right]
+\mathrm{h.c.}
\label{eq:expand}
\end{aligned}
\end{equation}

Substituting Eq.~\eqref{eq:int}~\eqref{eq:a_A} into Eq.~\eqref{eq:expand}, we obtain the  full form of the TCL2 master equation
\begin{equation}
\begin{aligned}
\frac{d\tilde{\rho}_S}{dt}
&=
\sum_{n,m}
e^{i(\omega_n-\omega_m)t}
\gamma_{\downarrow}(\omega_m,t)
\left(
A_m\tilde{\rho}_S A_n^\dagger
-
A_n^\dagger A_m\tilde{\rho}_S
\right)
\\
&\quad+
\sum_{n,m}
e^{-i(\omega_n-\omega_m)t}
\gamma_{\uparrow}(\omega_m,t)
\left(
A_m^\dagger\tilde{\rho}_S A_n
-
A_n A_m^\dagger\tilde{\rho}_S
\right)
+\mathrm{h.c.},
\end{aligned}
\end{equation}
where the time-dependent coefficients are determined by the bath correlation functions:
\begin{equation}
\gamma_{\downarrow}(\omega_n,t)
=
\int_0^t d\tau\,
e^{i\omega_n\tau}
\langle B(\tau)B^\dagger(0)\rangle,
\label{down}
\end{equation}
and
\begin{equation}
\gamma_{\uparrow}(\omega_n,t)
=
\int_0^t d\tau\,
e^{-i\omega_n\tau}
\langle B^\dagger(\tau)B(0)\rangle.
\label{up}
\end{equation}

We model the open-system dynamics by coupling the bosonic mode to a zero-temperature quantum bath and a classical noise field, with the total bath operator given by
\begin{equation}
B(t)=\xi(t)+g\sqrt{2\pi}\,b_{\mathrm{in}}(t)
=\xi(t)+\sqrt{\Gamma_1^\downarrow}\,b_{\mathrm{in}}(t),
\label{input_output}
\end{equation}
where $\xi(t)$ denotes the injected classical noise, $g=\sqrt{\Gamma_1^\downarrow/(2\pi)}$ is the coupling constant between the system and the continuous vacuum bath, with $\Gamma_1^\downarrow=1/\tau_S$, where $\tau_S$ is the lifetime of the bosonic mode. According to the Wiener--Khinchin theorem, the classical noise power spectral density $S(\omega)$ is given by the Fourier transform of the two-time correlation function~\cite{Clerk2010}:
\begin{equation}
S(\omega)
=
\int_{-\infty}^{+\infty}
d\tau\,
\langle\xi^*(\tau)\xi(0)\rangle
e^{i\omega\tau}.
\end{equation}
The second term in Eq.~\eqref{input_output} describes intrinsic single-photon loss, where
$
b_{\mathrm{in}}(t)
=
\frac{1}{\sqrt{2\pi}}
\int_{-\infty}^{+\infty}
d\omega\,b(\omega)e^{-i\omega t}
$
is the input field operator in standard input--output theory~\cite{Gardiner1985}.

The vacuum fluctuations are treated as a zero-temperature broadband white noise satisfying the delta-correlated statistics
$\langle b_{\mathrm{in}}(\tau)b_{\mathrm{in}}^\dagger(0)\rangle
=
\delta(\tau)$ and
$\langle b_{\mathrm{in}}^\dagger(\tau)b_{\mathrm{in}}(0)\rangle = 0$.

The resulting bath correlation functions are therefore
\begin{equation}
\langle B(\tau)B^\dagger(0)\rangle
=
\langle\xi(\tau)\xi^*(0)\rangle
+
\Gamma_1^\downarrow\delta(\tau),
\label{correlation1}
\end{equation}
and
\begin{equation}
\langle B^\dagger(\tau)B(0)\rangle
=
\langle\xi^*(\tau)\xi(0)\rangle.
\label{correlation2}
\end{equation}

Substituting Eqs.~\eqref{correlation1} and \eqref{correlation2} into Eqs.~\eqref{down} and \eqref{up}, respectively, the coefficients can be separated into their dissipative and dispersive contributions:
\begin{equation}
\gamma_{\downarrow}(\omega_n,t)
=
\frac{\tilde{\gamma}(\omega_n,t)+\Gamma_1^\downarrow}{2}
+iL(\omega_n,t),
\label{eq:gamma_d}
\end{equation}
and
\begin{equation}
\gamma_{\uparrow}(\omega_n,t)
=
\frac{\tilde{\gamma}(\omega_n,t)}{2}
-iL(\omega_n,t).
\label{eq:gamma_u}
\end{equation}
The quantity
$
L(\omega_n,t)
=
\int_{-\infty}^{+\infty}
d\omega'\,
S(\omega')
\frac{1-\cos[(\omega_n-\omega')t]}
{2\pi(\omega_n-\omega')}
$
is the time-dependent Lamb shift. Since we focus on the Fock-state populations, the Lamb-shift contribution can be omitted, as it does not affect the population dynamics.
The additional time-dependent dissipation rate induced by the classical noise is
$
\tilde{\gamma}(\omega_n,t)
=
\int_{-\infty}^{+\infty}
d\omega'\,
S(\omega')
\frac{\sin[(\omega'-\omega_n)t]}
{\pi(\omega'-\omega_n)}
\label{gamma_t}.
$
The expression for $\tilde{\gamma}(\omega_n,t)$ shows that, at a finite evolution time $t$, the system samples the noise spectrum through a sinc-shaped spectral filter with a characteristic width of order $1/t$.
For the classical noise considered here, the positive- and negative-frequency components are symmetric, $S(\omega)=S(-\omega)$, resulting in equal upward and downward transition rates. In the notation used in the main text,
\begin{equation}
\tilde{\gamma}_n^\downarrow(t)
=
\tilde{\gamma}_{n-1}^\uparrow(t)
=
\tilde{\gamma}(\omega_n,t).
\end{equation}

The resulting equation governing the diagonal dynamics can then be written as
\begin{equation}
\begin{aligned}
\frac{d\tilde{\rho}_S}{dt}\biggr\rvert_{\mathrm{diag}}
&=
\sum_{n=1}^{\infty}
\tilde{\gamma}_n^\downarrow(t)
\mathcal{D}[A_n]\tilde{\rho}_S
+
\sum_{n=1}^{\infty}
\tilde{\gamma}_{n-1}^\uparrow(t)
\mathcal{D}[A_n^\dagger]\tilde{\rho}_S
+
\Gamma_1^\downarrow\mathcal{D}[a]\tilde{\rho}_S .
\end{aligned}
\end{equation}

The intrinsic single-photon loss \(\Gamma_1^\downarrow\) does not distinguish between the individual Kerr-shifted transitions and is therefore described by the collective dissipator \(\mathcal{D}[a]\tilde{\rho}_S\), where $\mathcal{D}[a]\tilde{\rho}_S = a\tilde{\rho}_S a^\dagger -\frac{1}{2}\{a^\dagger a,\tilde{\rho}_S\}.$ By contrast, the structured noise spectrally resolves the Kerr-shifted transitions, giving rise to distinct, independently driven transition channels.

For the population dynamics, the transition between adjacent Fock states $\vert n\rangle$ and $\vert n-1\rangle$ acquires a bosonic enhancement factor $n$ originating from the matrix element of
$A_n=\sqrt{n}\vert n-1\rangle\langle n\vert$.
The resulting time-dependent transition rate is therefore
\begin{equation}
\tilde{\Gamma}_n^\downarrow(t)
=
\tilde{\Gamma}_{n-1}^\uparrow(t)
=
n\tilde{\gamma}(\omega_n,t)
=
n\int_{-\infty}^{+\infty}
d\omega'\,
S(\omega')
\frac{\sin[(\omega'-\omega_n)t]}
{\pi(\omega'-\omega_n)}.
\label{NM_rate}
\end{equation}

Equation~\eqref{NM_rate} explicitly captures the finite-time spectral response of the system. For $t\lesssim\tau_c$, the finite width of the sinc filter allows the transition rate to sample a finite frequency range of the structured noise spectrum, resulting in a time-dependent rate and transient non-Markovian dynamics. In the long-time limit, $t\gg\tau_c$, the sinc filter becomes sharply peaked around $\omega_n$ and approaches a delta function,
$
\frac{\sin[(\omega'-\omega_n)t]}
{\pi(\omega'-\omega_n)}
\longrightarrow
\delta(\omega'-\omega_n).
$
Consequently, the time-dependent transition rate converges to the Markovian rate
\begin{equation}
\tilde{\Gamma}_n^{\downarrow,L}
=
\tilde{\Gamma}_{n-1}^{\uparrow,L}
=
nS(\omega_n),
\label{M_rate}
\end{equation}
which exhibits the expected bosonic scaling with photon number and recovers the transition rate given by Fermi's golden rule~\cite{Clerk2010}.

\subsection{Transition rates under a Gaussian noise spectrum}
\label{Gaussian_Noise_Rates}

In our experiment, the injected classical noise is described by a symmetric two-sided Gaussian power spectral density,
\begin{equation}
S(\omega)
=
\eta
\left[
\exp\left(-\frac{(\omega-\omega_{\mathrm{c}})^2}{2\sigma^2}\right)
+
\exp\left(-\frac{(\omega+\omega_{\mathrm{c}})^2}{2\sigma^2}\right)
\right],
\label{eq:Gaussian_spectrum}
\end{equation}
where $\omega_{\mathrm{c}}$ is the center frequency of the noise, $\sigma$ denotes its spectral standard deviation, and $\eta$ is the peak spectral density. In the experiment, $\sigma/2\pi \simeq 5~\mathrm{kHz}$. The positive- and negative-frequency components are symmetric, $S(\omega)=S(-\omega)$, leading to equal noise-induced upward and downward transition rates between adjacent Fock states.

Substituting the Gaussian noise spectrum into Equation~(\ref{NM_rate}), the negative-frequency component centered at $-\omega_{\mathrm{c}}$ can be neglected for the transition near $+\omega_n$, since $\omega_{\mathrm{c}},\omega_n \gg \sigma$. Equation~(\ref{NM_rate}) therefore reduces to
\begin{equation}
\widetilde{\Gamma}^{\downarrow}_n(t)
\simeq
\frac{n\eta}{\pi}
\int_{-\infty}^{+\infty}d\omega'\,
\exp\left[-\frac{(\omega'-\omega_{\mathrm{c}})^2}{2\sigma^2}\right]
\frac{\sin[(\omega'-\omega_n)t]}
{\omega'-\omega_n}.
\label{eq:TCL2_single_Gaussian}
\end{equation}
Defining the detuning between the noise center and the transition frequency as
\begin{equation}
\Delta_n=\omega_{\mathrm{c}}-\omega_n,
\end{equation}
the integral can be evaluated analytically, yielding
\begin{equation}
\widetilde{\Gamma}^{\downarrow}_n(t)
=
n\eta
\exp\left(-\frac{\Delta_n^2}{2\sigma^2}\right)
\operatorname{Re}
\left[
\operatorname{erf}
\left(
\frac{\sigma t-i\Delta_n/\sigma}{\sqrt{2}}
\right)
\right],
\label{eq:TCL2_Gaussian_analytic}
\end{equation}
where $\operatorname{erf}(z)=(2/\sqrt{\pi})\int_0^z e^{-u^2}du$ is the error function.

For a resonant transition, $\Delta_n=0$, Eq.~(\ref{eq:TCL2_Gaussian_analytic}) reduces to
\begin{equation}
\widetilde{\Gamma}^{\downarrow}_n(t)
=
n\eta\,
\operatorname{erf}\left(\frac{\sigma t}{\sqrt{2}}\right),
\label{eq:resonant_rate}
\end{equation}
which increases monotonically from zero and approaches the Markovian rate $n\eta$ at long times. In contrast, for an off-resonant transition with $\Delta_n\neq0$, the finite detuning introduces transient oscillations in the TCL2 rate. The characteristic oscillation angular frequency is approximately set by $|\Delta_n|$, corresponding to an oscillation period
\begin{equation}
T_{\mathrm{osc}}
\simeq
\frac{2\pi}{|\Delta_n|}.
\label{eq:oscillation_period}
\end{equation}
Therefore, for the Kerr-resolved transitions of our system, whose adjacent transition frequencies are separated by the Kerr nonlinearity $K$, transitions detuned from the noise center by one and two Kerr spacings exhibit characteristic periods
\begin{equation}
T_{\Delta=K}\simeq\frac{2\pi}{K},
\qquad
T_{\Delta=2K}\simeq\frac{\pi}{K}.
\label{eq:Kerr_oscillation_period}
\end{equation}
This behavior is shown in Fig.~3(c) of the main text. The increasing downward deviation of the transition rates extracted from the Markovian fits at higher Fock-state numbers in Fig.~3(d) can be understood from the relative timescales of the noise correlation and the experimental observation window. Since the noise bandwidth is fixed, the noise correlation time $\tau_c$ remains approximately the same for different $m$. In contrast, due to bosonic enhancement of the intrinsic cavity decay, the intrinsic lifetime $T_m$ of the Fock state $\ket{m}$ decreases approximately as $1/m$. For each $m$, the population dynamics are measured over a time window of approximately $6T_m$. Consequently, as $m$ increases, the transient non-Markovian regime occupies an increasingly large fraction of the observation window. During this transient regime, the time-dependent TCL2 transition rate $\widetilde{\Gamma}^{\downarrow}_m(t)$ has not yet reached its long-time value. A Markovian fit with a constant transition rate to the entire decay curve is therefore increasingly influenced by the transient dynamics, resulting in the progressively larger downward deviation observed in Fig.~3(d).

In the long-time limit, the sinc kernel approaches a delta function,
\begin{equation}
\lim_{t\rightarrow\infty}
\frac{\sin[(\omega'-\omega_n)t]}
{\pi(\omega'-\omega_n)}
=
\delta(\omega'-\omega_n),
\end{equation}
and the TCL2 rate consequently converges to
\begin{equation}
\widetilde{\Gamma}^{\downarrow}_n(\infty)
=
nS(\omega_n)
\simeq
n\eta
\exp\left(-\frac{\Delta_n^2}{2\sigma^2}\right),
\label{eq:Gaussian_Markov_rate}
\end{equation}
recovering the Markovian transition rate given by Fermi's golden rule. Thus, the finite noise bandwidth determines both the transient non-Markovian behavior and the long-time frequency selectivity: $\sigma$ sets the correlation timescale, while the detuning $\Delta_n$ determines the characteristic transient oscillation frequency.

\section{Data Analysis and Numerical Simulations}

\subsection{TCL2 and Markovian Fitting Procedures}
\label{Fitting_Details}

For both the TCL2 and Markovian fits, we use the Gaussian noise spectrum in Eq.~\eqref{eq:Gaussian_spectrum}, with the center frequency fixed at $\omega_{\mathrm{LO}}$ and the spectral standard deviation fixed at $\sigma/2\pi=5.7~\mathrm{kHz}$. The noise intensity $\eta$ is the only fitting parameter. The intrinsic cavity decay is included using the independently measured single-photon decay rate $\Gamma_1^{\downarrow}$.

For the TCL2 model, the time-dependent noise-induced transition rates $\widetilde{\Gamma}_{n}^{\uparrow,\downarrow}(t)$ are calculated from Eq.~\eqref{eq:TCL2_Gaussian_analytic}. The population dynamics are obtained by integrating
\begin{equation}
\begin{aligned}
\frac{dP_n(t)}{dt}
=
&\widetilde{\Gamma}_{n+1}^{\downarrow}(t)P_{n+1}(t)
+\widetilde{\Gamma}_{n-1}^{\uparrow}(t)P_{n-1}(t) \\
&-\left[
\widetilde{\Gamma}_{n}^{\downarrow}(t)
+\widetilde{\Gamma}_{n}^{\uparrow}(t)
\right]P_n(t) \\
&+(n+1)\Gamma_1^{\downarrow}P_{n+1}(t)
-n\Gamma_1^{\downarrow}P_n(t).
\end{aligned}
\label{eq:TCL2_population_rate}
\end{equation}
For the symmetric classical noise spectrum considered here, $S(\omega)=S(-\omega)$, the upward and downward rates associated with the same transition satisfy
$
\widetilde{\Gamma}_{n-1}^{\uparrow}(t)
=
\widetilde{\Gamma}_{n}^{\downarrow}(t).
$
For each trial value of $\eta$, Eq.~\eqref{eq:TCL2_population_rate} is numerically integrated to obtain the theoretical populations $P_n(t)$ for fitting the experimental data.

For comparison, the Markovian model uses time-independent noise-induced transition rates,
$
\widetilde{\Gamma}_{n}^{\downarrow,\mathrm{M}}
=
nS(\omega_n),
$
with
$
\widetilde{\Gamma}_{n-1}^{\uparrow,\mathrm{M}}
=
\widetilde{\Gamma}_{n}^{\downarrow,\mathrm{M}}
$
for the symmetric noise spectrum. The corresponding population dynamics are obtained from
\begin{equation}
\begin{aligned}
\frac{dP_n(t)}{dt}
=
&\widetilde{\Gamma}_{n+1}^{\downarrow,\mathrm{M}}P_{n+1}(t)
+\widetilde{\Gamma}_{n-1}^{\uparrow,\mathrm{M}}P_{n-1}(t) \\
&-\left[
\widetilde{\Gamma}_{n}^{\downarrow,\mathrm{M}}
+\widetilde{\Gamma}_{n}^{\uparrow,\mathrm{M}}
\right]P_n(t) \\
&+(n+1)\Gamma_1^{\downarrow}P_{n+1}(t)
-n\Gamma_1^{\downarrow}P_n(t).
\end{aligned}
\label{eq:Markov_population_rate}
\end{equation}
For each trial value of $\eta$, Eq.~\eqref{eq:Markov_population_rate} is integrated to obtain $P_n(t)$ for fitting the experimental data. Thus, the TCL2 and Markovian models use the same noise spectrum, intrinsic decay, and fitting parameter, differing only in whether the noise-induced transition rates are time dependent or time independent. 

Since the spectral fit in Fig.~2(d) of the main text shows a $0.33~\mathrm{kHz}$ offset of the noise center frequency from zero detuning, this offset is included in the fitting. 
For fitting the experimental data, the populations calculated using either the TCL2 or Markovian model are mapped onto the measured signals according to
\begin{equation}
P_n^{\mathrm{fit}}(t)
=
A P_n(t)+C,
\label{eq:population_mapping}
\end{equation}
where $A$ is the measurement response amplitude and $C$ is the baseline offset. The initial population distribution $P(0)$ is obtained by normalizing the measured signals at $t=0$, while $\eta$, $A$, and $C$ are treated as fitting parameters.

\begin{figure*}[htbp]
    \centering
    \includegraphics{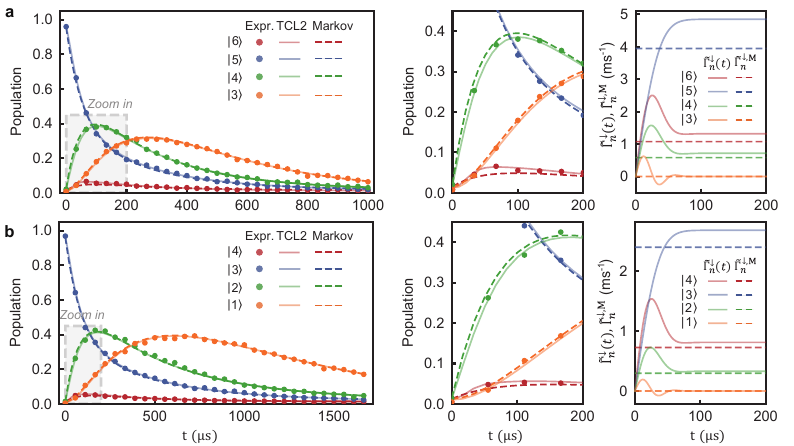}
   \caption{\textbf{Population dynamics for initial Fock states $|5\rangle$ and $|3\rangle$.}
Experimental population dynamics (dots) and corresponding TCL2 and Markovian fits (lines) for the initial states $|m\rangle=|5\rangle$ (\textbf{a}) and $|m\rangle=|3\rangle$ (\textbf{b}).}
    \label{fig:S_fitting}
\end{figure*}

While the main text focuses on the initial state $|m\rangle=|10\rangle$, Fig.~\ref{fig:S_fitting} shows the fitting results for $|m\rangle=|3\rangle$ and $|5\rangle$. For smaller $m$, both the TCL2 and Markovian models agree well with the experiment, whereas their difference becomes more pronounced with increasing $m$, indicating stronger non-Markovian effects, consistent with Fig.~3(d) of the main text.

\subsection{Fitting with the noise center frequency as a free parameter $\omega_{\mathrm{c}}$}

In the main-text fits, the noise center frequency is fixed at the experimentally set $\omega_{\mathrm{LO}}$, with the measured $0.33~\mathrm{kHz}$ offset subtracted. Here, we allow the noise center frequency $\omega_{\mathrm{c}}$ to vary as an additional free parameter to examine its effect on the fitting.

\begin{figure*}[htbp]
\centering
\includegraphics{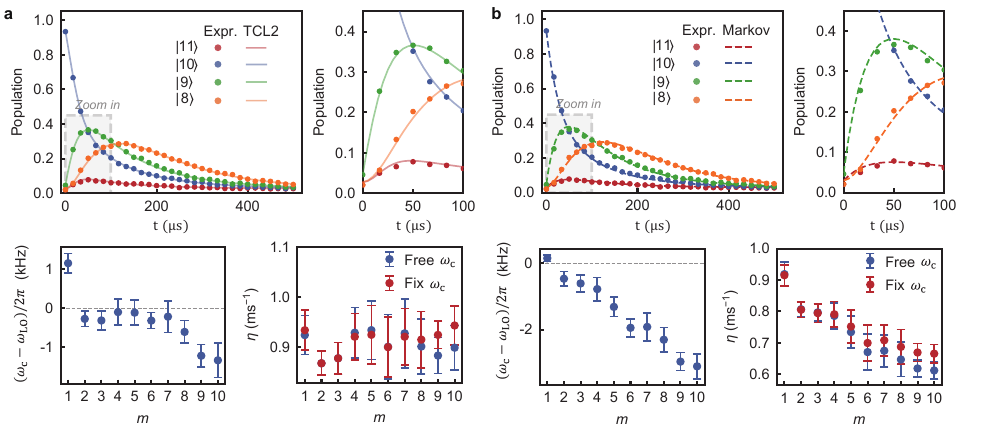}
\caption{\textbf{Fitting with the noise center frequency $\omega_{\mathrm{c}}$ as a free parameter.}
\textbf{a} TCL2 fitting results. The upper panels show the population dynamics for $\ket{m}=\ket{10}$, with experimental data (dots) and fits (solid lines); the right panel enlarges the early-time dynamics. The lower panels show the fitted frequency offset $\omega_{\mathrm{c}}-\omega_{\mathrm{LO}}$ (with the $0.33~\mathrm{kHz}$ shift subtracted) and noise strength $\eta$ for different initial Fock states $\ket{m}$. Blue and red denote $\eta$ obtained with $\omega_{\mathrm{c}}$ free and fixed at $\omega_{\mathrm{LO}}$, respectively. \textbf{b} Corresponding results for the Markovian model, with fits shown as dashed lines. Error bars represent the standard deviations of repeated experiments.}
\label{fig:w0fit}
\end{figure*}

As shown in Fig.~\ref{fig:w0fit}, varying $\omega_{\mathrm{c}}$ only slightly improves the TCL2 fit but noticeably improves the Markovian fit. For $\ket{m}=\ket{10}$, the root-mean-square error (RMSE), averaged over the nine measured Fock-state populations, decreases from $0.0050$ to $0.0048$ for the TCL2 model and from $0.0088$ to $0.0073$ for the Markovian model.

The improved fits, however, require systematic shifts of $\omega_{\mathrm{c}}$ from $\omega_{\mathrm{LO}}$. The shift increases with $m$, reaching approximately $1.5~\mathrm{kHz}$ for the TCL2 model and nearly $3~\mathrm{kHz}$ for the Markovian model at $\ket{m}=\ket{10}$ [Fig.~\ref{fig:w0fit}]. In contrast, the fitted noise strength $\eta$ remains close to that obtained with $\omega_{\mathrm{c}}$ fixed and to the independently calibrated value of approximately $0.94~\mathrm{ms}^{-1}$.

These frequency shifts far exceed the uncertainty of the independently calibrated Kerr nonlinearity ($\sim0.02~\mathrm{kHz}$; see Sec.~\ref{sec:kerr_cal}). Thus, allowing $\omega_{\mathrm{c}}$ to vary provides little improvement for the TCL2 model, while the improved Markovian fit requires substantial frequency shifts inconsistent with the experimental calibration, indicating overfitting. We therefore fix the noise center frequency at $\omega_{\mathrm{LO}}$ in all main-text fits.

\subsection{Numerical Simulations}

We perform direct stochastic simulations of the system dynamics. Setting $\hbar=1$, the system is described by
\begin{equation}
H=
\omega_{\mathrm{S}}a^\dagger a
-\frac{K}{2}a^\dagger a^\dagger aa
+\xi(t)(a^\dagger+a),
\label{sim}
\end{equation}
where $\omega_{\mathrm{S}}$ and $K$ are the bosonic-mode frequency and Kerr nonlinearity, respectively, and $\xi(t)$ is the stochastic drive generated from the noise power spectrum $S(\omega)$.

We generate $\xi(t)$ using spectral synthesis~\cite{shinozuka1972}. Discretizing the two-sided spectrum $S(\omega)$ on a frequency grid $\omega_k$ with spacing $\delta\omega$, the stochastic waveform is
\begin{equation}
\xi(t)=
\sum_k
\sqrt{\frac{S(\omega_k)\delta\omega}{\pi}}
\cos(\omega_k t+\phi_k),
\label{eq:spectral_synthesis}
\end{equation}
where $\phi_k$ are independent random phases uniformly distributed over $[0,2\pi)$. Ensemble averaging gives
\begin{equation}
\left\langle\xi(\tau)\xi(0)\right\rangle
=
\frac{1}{2\pi}
\sum_k S(\omega_k)\cos(\omega_k\tau)\delta\omega
\rightarrow
\frac{1}{2\pi}
\int_{-\infty}^{+\infty}
S(\omega)\cos(\omega\tau)\,d\omega,
\end{equation}
which, by the Wiener--Khinchin theorem~\cite{Clerk2010}, ensures that $\xi(t)$ reproduces the prescribed spectrum $S(\omega)$.

The dynamics are obtained by numerically integrating the Lindblad master equation, including intrinsic single-photon loss $\Gamma_1^\downarrow\mathcal{D}[a]$. The Hilbert space is truncated at $N_{\max}=15$, and the Fock-state populations $P_n(t)$ are averaged over 500 independent noise realizations.

\begin{figure}[htbp]
    \centering
    \includegraphics{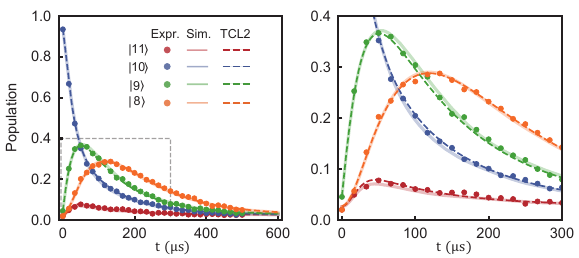}
    \caption{\textbf{Direct stochastic simulation of the non-Markovian dynamics.}
    Population dynamics of the Fock states $\ket{8}$--$\ket{11}$ corresponding to Fig.~3(b) of the main text. Dots show the experimental data, solid lines the direct stochastic simulations, and dashed lines the TCL2 fits. The right panel enlarges the first $300~\mu\mathrm{s}$.}
    \label{fig:sm_fig3}
\end{figure}

Figure~\ref{fig:sm_fig3} compares the direct stochastic simulations with the experimental dynamics in Fig.~3(b) of the main text. The simulations reproduce the measured population dynamics, including the early-time non-Markovian behavior.

\section{Experimental Details}
\label{sec:exp_details}
\subsection{Fock-State Preparation}

\begin{figure}[htbp]
    \centering
    \includegraphics{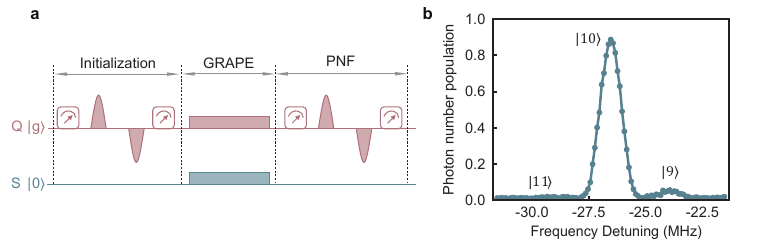}
   \caption{\textbf{Fock-state preparation and characterization.}
\textbf{a} Pulse sequence for Fock-state preparation using GRAPE followed by a heralded photon-number filter (PNF).
\textbf{b} Photon-number-splitting spectrum of the prepared Fock state $|10\rangle$.}
    \label{fig:sequence}
\end{figure}

The initial Fock state $|m\rangle$ is prepared using gradient ascent pulse engineering (GRAPE)~\cite{Khaneja2005}, followed by a heralded photon-number filter (PNF)~\cite{deng2024} to further enhance the state-preparation fidelity. The PNF employs a dispersive phase $\theta=\pi$ to suppress the dominant residual populations in the neighboring Fock states $|m\pm1\rangle$. Figure~\ref{fig:sequence} shows the preparation sequence and the photon-number-splitting spectrum of $|10\rangle$.

\subsection{Cavity Kerr Calibration}
\label{sec:kerr_cal}
\begin{figure}[htbp]
    \centering
\includegraphics{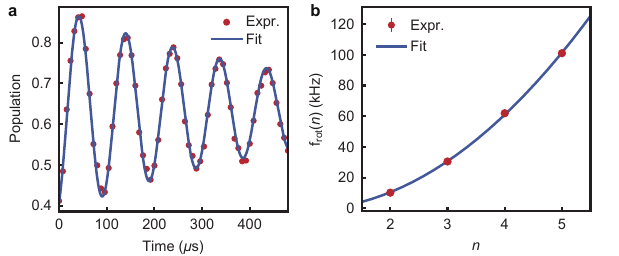}
\caption{\textbf{Calibration of the cavity Kerr nonlinearity.}
\textbf{a} Ramsey oscillations of the superposition state $(|0\rangle+|2\rangle)/\sqrt{2}$, measured in a round-robin sequence (8,000 shots per measurement). Red dots show the experimental data and the blue solid line is the fit.
\textbf{b} Extracted rotation frequency $f_{\mathrm{rot}}(n)$ for different Fock-state superpositions $(|0\rangle+|n\rangle)/\sqrt{2}$. Each point is averaged over 4 independent experiments; error bars are smaller than the symbols. The solid line is a fit to $f_{\mathrm{rot}}(n)$. The four independent fits yield $K/2\pi=10.1\pm0.02~\mathrm{kHz}$, where the uncertainty is the standard deviation of the fitted Kerr values.}
    \label{fig:ker_0pN}
\end{figure}

To precisely calibrate the cavity Kerr nonlinearity~\cite{Kirchmair2013}, we prepare a series of Fock-state superpositions $(\ket{0}+\ket{n})/\sqrt{2}$ with $n=2,3,4,5$, and measure their relative phase evolution as a function of time (Fig.~\ref{fig:ker_0pN}a). For the cavity Hamiltonian $H_S/\hbar=\omega_S\hat{n}-K\hat{n}(\hat{n}-1)/2$, the Kerr-induced energy shift scales as $n(n-1)$, resulting in an $n$-dependent relative phase accumulation between the $\ket{0}$ and $\ket{n}$ components. In the rotating frame, the corresponding rotation frequency is $f_{\mathrm{rot}}(n)=-\frac{K}{4\pi}n(n-1)+\delta_{\omega_S}$, where $\delta_{\omega_S}$ accounts for an $n$-independent frequency offset. We extract $f_{\mathrm{rot}}(n)$ from the measured phase evolution of each superposition state and fit its photon-number dependence (Fig.~\ref{fig:ker_0pN}b). The cavity Kerr nonlinearity is $K/2\pi=10.1\pm0.02$ kHz.

\subsection{Noise Spectrum Measurement with a Spectrum Analyzer}

\begin{figure}[htbp]
    \centering
    \includegraphics{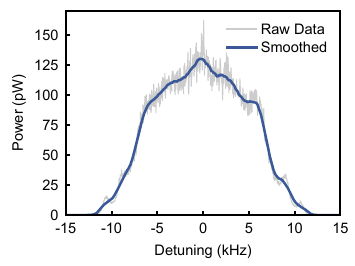}
    \caption{\textbf{Measurement of the noise power spectrum.} The light gray line shows the noise spectrum averaged over $N=30$ independent traces, and the blue curve shows the smoothed noise envelope obtained using a 50-point moving average.}
    \label{fig:noise_spectrum}
\end{figure}

The noise power spectrum was measured using a spectrum analyzer centered at $\omega_S$ with a 30~kHz frequency span and 1000 sweep points. The sample detector was used, with the resolution bandwidth (RBW) and video bandwidth (VBW) set to 100~Hz and 10~Hz, respectively. We averaged $N=30$ independent traces to reduce statistical fluctuations and subsequently applied a 50-point moving average to obtain the smoothed noise envelope shown in Fig.~\ref{fig:noise_spectrum}. In Fig.~2(d) of the main text, the effective noise spectrum is obtained by sweeping $\omega_{\mathrm{LO}}$ across $\omega_S$. The system therefore probes the noise spectrum $S(\omega)$ at $\omega=\omega_S-\omega_{\mathrm{LO}}$, such that increasing $\omega_{\mathrm{LO}}$ corresponds to sampling $S(\omega)$ from positive to negative frequencies. The frequency axis of the spectrum-analyzer data is therefore reversed in Fig.~2(d) for direct comparison.

\subsection{Extraction of Intrinsic Decay Rates $\Gamma_n^\downarrow$}

To determine the intrinsic decay rates $\Gamma_n^\downarrow$ of the Fock states, we employ a sequential cascade decay model~\cite{Wang2008,Wang2025}. For an initial Fock state $|n\rangle$, the decay dynamics are described by cascaded transitions between adjacent Fock states. The time-dependent populations $P_j(t)$ are obtained by integrating the following rate equations:
\begin{equation}
\frac{dP_j(t)}{dt}
=
\Gamma_{j+1}^{\downarrow}P_{j+1}(t)
-
\Gamma_j^{\downarrow}P_j(t).
\label{eq:cascade_population_rate}
\end{equation}
The calculated populations are mapped onto the measured signals as $A P_j(t)+C$.

\begin{figure}[htbp]
    \centering
    \includegraphics{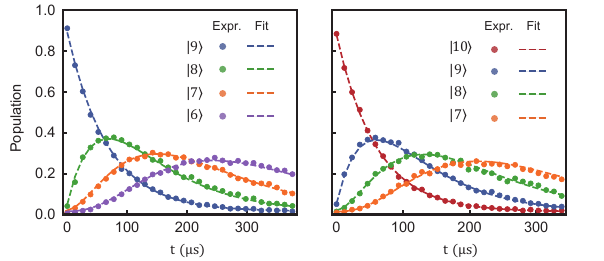}
   \caption{\textbf{Intrinsic Fock-state decay rates from cascade dynamics.}
Population dynamics for the initial Fock states $|9\rangle$ (\textbf{left}) and $|10\rangle$ (\textbf{right}). The markers show the experimental data, while the dashed lines represent fits based on the cascade decay model in Eq.~\ref{eq:cascade_population_rate}, with the previously determined lower-Fock-state decay rates fixed.}
    \label{fig:fock_cascade}
\end{figure}

Experimentally, we prepare the Fock states from $|1\rangle$ to $|10\rangle$ separately and determine their decay rates sequentially from low to high Fock states. We first prepare $|1\rangle$ and fit its population decay to obtain $\Gamma_1^\downarrow$. We then prepare $|2\rangle$ and determine $\Gamma_2^\downarrow$ with the previously extracted $\Gamma_1^\downarrow$ fixed. For higher Fock states, this procedure is repeated: for each prepared state $|n\rangle$, we measure the population dynamics from $|n\rangle$ to $|n-3\rangle$ and fit $\Gamma_n^\downarrow$ while fixing all previously determined decay rates $\Gamma_1^\downarrow,\ldots,\Gamma_{n-1}^\downarrow$ (Fig.~\ref{fig:fock_cascade}). By iterating this procedure from $|1\rangle$ to $|10\rangle$, we obtain the intrinsic decay rates of all measured Fock states. A linear fit to the data in Fig.~2(c) of the main text, $\Gamma_n^\downarrow=\Gamma_1^\downarrow n+b$, yields $\Gamma_1^\downarrow\approx1.66\pm0.003~\mathrm{ms}^{-1}$ and $b=-0.0641\pm0.05~\mathrm{ms}^{-1}$. The near-zero intercept is consistent with the expected bosonic scaling $\Gamma_n^\downarrow=n\Gamma_1^\downarrow$. The extracted decay rate, corresponding to \(T_{1S}\approx602~\mu\mathrm{s}\), is used as the intrinsic decay-rate coefficient in all subsequent fits.

\subsection{Complete Fock-State Population Dynamics of the Bosonic Logical Qubit}

\begin{figure}[htbp]
    \centering
    \includegraphics{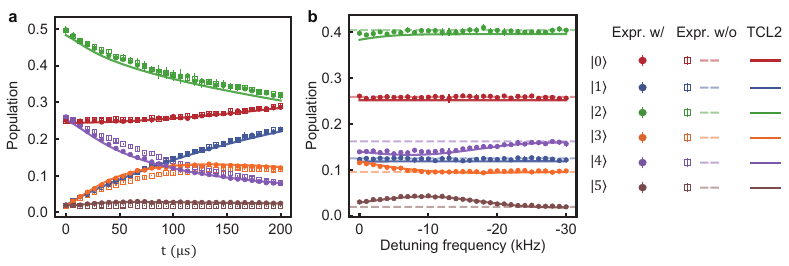}
    \caption{\textbf{Full population dynamics of the lowest-order binomial code.}
\textbf{a} Time evolution of the Fock-state populations from $|0\rangle$ to $|5\rangle$ after preparing $|+\rangle_{\mathrm L}$. Filled circles and open squares denote the dynamics with and without narrowband noise centered at $\omega_4$, respectively. Solid lines are the TCL2 predictions.
\textbf{b} Fock-state populations at $t=75\,\mu\mathrm{s}$ as a function of the noise center frequency. Dashed lines indicate the corresponding values without noise.}
    \label{fig:S_binominal}
\end{figure}

In the main text (Fig.~4), we focus on the error states $|1\rangle$ and $|3\rangle$ to illustrate how the engineered narrowband noise modifies the error dynamics of the bosonic logical qubit. To provide a more complete physical picture, Supplementary Fig.~\ref{fig:S_binominal} shows the population dynamics of all relevant Fock states from $|0\rangle$ to $|5\rangle$ for the lowest-order binomial code. To account for state-preparation imperfections, the experimentally measured initial-state populations are directly used in the theoretical model.

As shown in Fig.~\ref{fig:S_binominal}a, when the noise is centered at $\omega_4$, the increase in the $|3\rangle$ population observed in the main text originates predominantly from the noise-induced decay of the $|4\rangle$ component, while the populations of the other Fock states remain nearly unchanged. This directly demonstrates the Fock-state selectivity of the engineered narrowband noise. The TCL2 predictions are in good agreement with the measured population dynamics.
Figure~\ref{fig:S_binominal}b further demonstrates this frequency selectivity. As the noise center frequency is tuned from $\omega_4$ toward the neighboring $\omega_5$, the populations of $|4\rangle$ and $|5\rangle$ are increasingly affected.
As the noise is further detuned from these transitions, the noise-induced effects are suppressed and the populations of all Fock states recover to their values in the absence of injected noise.

\subsection{System Dynamics under Narrowband Noise with a 1-kHz Low-Pass Filter}
\label{sec:2k_results}

To further investigate the system dynamics with stronger non-Markovian effects, we repeat the experiment in Fig.~3(b) of the main text under modified noise conditions. Specifically, we use a 1~kHz instead of a 5~kHz low-pass filter and remove one 6~dB attenuator. The system is initially prepared in the Fock state $|10\rangle$, with the noise centered at $\omega_9$. The narrower bandwidth increases the noise correlation time, while the higher noise strength enhances the induced transitions.

\begin{figure}[htbp]
    \centering
    \includegraphics{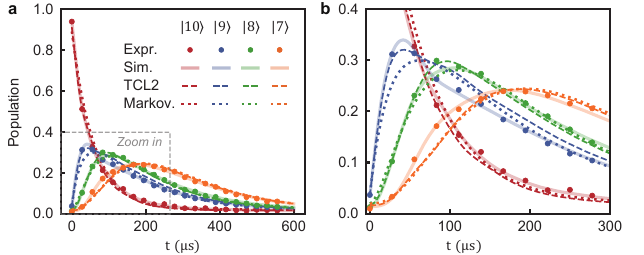}
\caption{\textbf{System dynamics under narrower-band noise.}
The system is initially prepared in $|10\rangle$, with the noise centered at $\omega_9$. Compared with Fig.~3(b) of the main text, we use a 1~kHz instead of a 5~kHz low-pass filter and remove one 6~dB attenuator. \textbf{a} Time evolution of the Fock-state populations. Dots show the experimental data, Solid lines show direct stochastic simulation results based on spectral synthesis; dashed and dotted lines show the TCL2 and Markovian fits, respectively. \textbf{b} Zoom-in of the early-time region indicated by the gray dashed box in \textbf{a}. The narrower-band and stronger noise leads to more pronounced non-Markovian dynamics. The TCL2 fit captures the early-time dynamics but gradually deviates at longer times, whereas the direct stochastic simulation agrees well with the experiment throughout the evolution.}
\label{fig:res_2k}
\end{figure}

As shown in Fig.~\ref{fig:res_2k}a, the measured population dynamics exhibit more pronounced non-Markovian behavior than in Fig.~3(b) of the main text, with the early-time region enlarged in Fig.~\ref{fig:res_2k}b. The TCL2 fit captures the early-time dynamics but gradually deviates at longer times, reflecting the limitation of the second-order approximation under the longer-correlated and stronger noise. In contrast, the direct stochastic simulation based on spectral synthesis reproduces the experimental dynamics throughout the evolution.

\begin{thebibliography}{49}%
\makeatletter
\providecommand \@ifxundefined [1]{%
 \@ifx{#1\undefined}
}%
\providecommand \@ifnum [1]{%
 \ifnum #1\expandafter \@firstoftwo
 \else \expandafter \@secondoftwo
 \fi
}%
\providecommand \@ifx [1]{%
 \ifx #1\expandafter \@firstoftwo
 \else \expandafter \@secondoftwo
 \fi
}%
\providecommand \natexlab [1]{#1}%
\providecommand \enquote  [1]{``#1''}%
\providecommand \bibnamefont  [1]{#1}%
\providecommand \bibfnamefont [1]{#1}%
\providecommand \citenamefont [1]{#1}%
\providecommand \href@noop [0]{\@secondoftwo}%
\providecommand \href [0]{\begingroup \@sanitize@url \@href}%
\providecommand \@href[1]{\@@startlink{#1}\@@href}%
\providecommand \@@href[1]{\endgroup#1\@@endlink}%
\providecommand \@sanitize@url [0]{\catcode `\\12\catcode `\$12\catcode `\&12\catcode `\#12\catcode `\^12\catcode `\_12\catcode `\%12\relax}%
\providecommand \@@startlink[1]{}%
\providecommand \@@endlink[0]{}%
\providecommand \url  [0]{\begingroup\@sanitize@url \@url }%
\providecommand \@url [1]{\endgroup\@href {#1}{\urlprefix }}%
\providecommand \urlprefix  [0]{URL }%
\providecommand \Eprint [0]{\href }%
\providecommand \doibase [0]{http://dx.doi.org/}%
\providecommand \selectlanguage [0]{\@gobble}%
\providecommand \bibinfo  [0]{\@secondoftwo}%
\providecommand \bibfield  [0]{\@secondoftwo}%
\providecommand \translation [1]{[#1]}%
\providecommand \BibitemOpen [0]{}%
\providecommand \bibitemStop [0]{}%
\providecommand \bibitemNoStop [0]{.\EOS\space}%
\providecommand \EOS [0]{\spacefactor3000\relax}%
\providecommand \BibitemShut  [1]{\csname bibitem#1\endcsname}%
\let\auto@bib@innerbib\@empty
\bibitem [{\citenamefont {Braunstein}\ and\ \citenamefont {van Loock}(2005)}]{Braunstein2005}%
  \BibitemOpen
  \bibfield  {author} {\bibinfo {author} {\bibfnamefont {S.~L.}\ \bibnamefont {Braunstein}}\ and\ \bibinfo {author} {\bibfnamefont {P.}~\bibnamefont {van Loock}},\ }\bibfield  {title} {\enquote {\bibinfo {title} {Quantum information with continuous variables},}\ }\href {\doibase 10.1103/RevModPhys.77.513} {\bibfield  {journal} {\bibinfo  {journal} {Rev. Mod. Phys.}\ }\textbf {\bibinfo {volume} {77}},\ \bibinfo {pages} {513} (\bibinfo {year} {2005})}\BibitemShut {NoStop}%
\bibitem [{\citenamefont {Ma}\ \emph {et~al.}(2021)\citenamefont {Ma}, \citenamefont {Puri}, \citenamefont {Schoelkopf}, \citenamefont {Devoret}, \citenamefont {Girvin},\ and\ \citenamefont {Jiang}}]{Ma2021}%
  \BibitemOpen
  \bibfield  {author} {\bibinfo {author} {\bibfnamefont {W.-L.}\ \bibnamefont {Ma}}, \bibinfo {author} {\bibfnamefont {S.}~\bibnamefont {Puri}}, \bibinfo {author} {\bibfnamefont {R.~J.}\ \bibnamefont {Schoelkopf}}, \bibinfo {author} {\bibfnamefont {M.~H.}\ \bibnamefont {Devoret}}, \bibinfo {author} {\bibfnamefont {S.~M.}\ \bibnamefont {Girvin}}, \ and\ \bibinfo {author} {\bibfnamefont {L.}~\bibnamefont {Jiang}},\ }\bibfield  {title} {\enquote {\bibinfo {title} {Quantum control of bosonic modes with superconducting circuits},}\ }\href {\doibase 10.1016/j.scib.2021.05.024} {\bibfield  {journal} {\bibinfo  {journal} {Sci. Bull.}\ }\textbf {\bibinfo {volume} {66}},\ \bibinfo {pages} {1789} (\bibinfo {year} {2021})}\BibitemShut {NoStop}%
\bibitem [{\citenamefont {Cai}\ \emph {et~al.}(2021)\citenamefont {Cai}, \citenamefont {Ma}, \citenamefont {Wang}, \citenamefont {Zou},\ and\ \citenamefont {Sun}}]{Cai2021}%
  \BibitemOpen
  \bibfield  {author} {\bibinfo {author} {\bibfnamefont {W.}~\bibnamefont {Cai}}, \bibinfo {author} {\bibfnamefont {Y.}~\bibnamefont {Ma}}, \bibinfo {author} {\bibfnamefont {W.}~\bibnamefont {Wang}}, \bibinfo {author} {\bibfnamefont {C.-L.}\ \bibnamefont {Zou}}, \ and\ \bibinfo {author} {\bibfnamefont {L.}~\bibnamefont {Sun}},\ }\bibfield  {title} {\enquote {\bibinfo {title} {Bosonic quantum error correction codes in superconducting quantum circuits},}\ }\href {\doibase 10.1016/j.fmre.2020.12.006} {\bibfield  {journal} {\bibinfo  {journal} {Fundam. Res.}\ }\textbf {\bibinfo {volume} {1}},\ \bibinfo {pages} {50} (\bibinfo {year} {2021})}\BibitemShut {NoStop}%
\bibitem [{\citenamefont {Joshi}\ \emph {et~al.}(2021)\citenamefont {Joshi}, \citenamefont {Noh},\ and\ \citenamefont {Gao}}]{Joshi2021}%
  \BibitemOpen
  \bibfield  {author} {\bibinfo {author} {\bibfnamefont {A.}~\bibnamefont {Joshi}}, \bibinfo {author} {\bibfnamefont {K.}~\bibnamefont {Noh}}, \ and\ \bibinfo {author} {\bibfnamefont {Y.~Y.}\ \bibnamefont {Gao}},\ }\bibfield  {title} {\enquote {\bibinfo {title} {Quantum information processing with bosonic qubits in circuit {QED}},}\ }\href {\doibase 10.1088/2058-9565/abe989} {\bibfield  {journal} {\bibinfo  {journal} {Quantum Sci. Technol.}\ }\textbf {\bibinfo {volume} {6}},\ \bibinfo {pages} {033001} (\bibinfo {year} {2021})}\BibitemShut {NoStop}%
\bibitem [{\citenamefont {Michael}\ \emph {et~al.}(2016)\citenamefont {Michael}, \citenamefont {Silveri}, \citenamefont {Brierley}, \citenamefont {Albert}, \citenamefont {Salmilehto}, \citenamefont {Jiang},\ and\ \citenamefont {Girvin}}]{Michael2016}%
  \BibitemOpen
  \bibfield  {author} {\bibinfo {author} {\bibfnamefont {M.~H.}\ \bibnamefont {Michael}}, \bibinfo {author} {\bibfnamefont {M.}~\bibnamefont {Silveri}}, \bibinfo {author} {\bibfnamefont {R.~T.}\ \bibnamefont {Brierley}}, \bibinfo {author} {\bibfnamefont {V.~V.}\ \bibnamefont {Albert}}, \bibinfo {author} {\bibfnamefont {J.}~\bibnamefont {Salmilehto}}, \bibinfo {author} {\bibfnamefont {L.}~\bibnamefont {Jiang}}, \ and\ \bibinfo {author} {\bibfnamefont {S.~M.}\ \bibnamefont {Girvin}},\ }\bibfield  {title} {\enquote {\bibinfo {title} {New class of quantum error-correcting codes for a bosonic mode},}\ }\href {\doibase 10.1103/PhysRevX.6.031006} {\bibfield  {journal} {\bibinfo  {journal} {Phys. Rev. X}\ }\textbf {\bibinfo {volume} {6}},\ \bibinfo {pages} {031006} (\bibinfo {year} {2016})}\BibitemShut {NoStop}%
\bibitem [{\citenamefont {Hu}\ \emph {et~al.}(2019)\citenamefont {Hu}, \citenamefont {Ma}, \citenamefont {Cai}, \citenamefont {Mu}, \citenamefont {Xu}, \citenamefont {Wang}, \citenamefont {Wu}, \citenamefont {Wang}, \citenamefont {Song}, \citenamefont {Zou}, \citenamefont {Girvin}, \citenamefont {Duan},\ and\ \citenamefont {Sun}}]{Hu2019}%
  \BibitemOpen
  \bibfield  {author} {\bibinfo {author} {\bibfnamefont {L.}~\bibnamefont {Hu}}, \bibinfo {author} {\bibfnamefont {Y.}~\bibnamefont {Ma}}, \bibinfo {author} {\bibfnamefont {W.}~\bibnamefont {Cai}}, \bibinfo {author} {\bibfnamefont {X.}~\bibnamefont {Mu}}, \bibinfo {author} {\bibfnamefont {Y.}~\bibnamefont {Xu}}, \bibinfo {author} {\bibfnamefont {W.}~\bibnamefont {Wang}}, \bibinfo {author} {\bibfnamefont {Y.}~\bibnamefont {Wu}}, \bibinfo {author} {\bibfnamefont {H.}~\bibnamefont {Wang}}, \bibinfo {author} {\bibfnamefont {Y.~P.}\ \bibnamefont {Song}}, \bibinfo {author} {\bibfnamefont {C.-L.}\ \bibnamefont {Zou}}, \bibinfo {author} {\bibfnamefont {S.~M.}\ \bibnamefont {Girvin}}, \bibinfo {author} {\bibfnamefont {L.-M.}\ \bibnamefont {Duan}}, \ and\ \bibinfo {author} {\bibfnamefont {L.}~\bibnamefont {Sun}},\ }\bibfield  {title} {\enquote {\bibinfo {title} {Quantum error correction and universal gate set operation on a binomial bosonic logical qubit},}\ }\href {\doibase 10.1038/s41567-018-0414-3} {\bibfield
  {journal} {\bibinfo  {journal} {Nat. Phys.}\ }\textbf {\bibinfo {volume} {15}},\ \bibinfo {pages} {503} (\bibinfo {year} {2019})}\BibitemShut {NoStop}%
\bibitem [{\citenamefont {Ma}\ \emph {et~al.}(2020)\citenamefont {Ma}, \citenamefont {Xu}, \citenamefont {Mu}, \citenamefont {Cai}, \citenamefont {Hu}, \citenamefont {Wang}, \citenamefont {Pan}, \citenamefont {Wang}, \citenamefont {Song}, \citenamefont {Zou},\ and\ \citenamefont {Sun}}]{Ma2020}%
  \BibitemOpen
  \bibfield  {author} {\bibinfo {author} {\bibfnamefont {Y.}~\bibnamefont {Ma}}, \bibinfo {author} {\bibfnamefont {Y.}~\bibnamefont {Xu}}, \bibinfo {author} {\bibfnamefont {X.}~\bibnamefont {Mu}}, \bibinfo {author} {\bibfnamefont {W.}~\bibnamefont {Cai}}, \bibinfo {author} {\bibfnamefont {L.}~\bibnamefont {Hu}}, \bibinfo {author} {\bibfnamefont {W.}~\bibnamefont {Wang}}, \bibinfo {author} {\bibfnamefont {X.}~\bibnamefont {Pan}}, \bibinfo {author} {\bibfnamefont {H.}~\bibnamefont {Wang}}, \bibinfo {author} {\bibfnamefont {Y.}~\bibnamefont {Song}}, \bibinfo {author} {\bibfnamefont {C.-L.}\ \bibnamefont {Zou}}, \ and\ \bibinfo {author} {\bibfnamefont {L.}~\bibnamefont {Sun}},\ }\bibfield  {title} {\enquote {\bibinfo {title} {Error-transparent operations on a logical qubit protected by quantum error correction},}\ }\href {\doibase 10.1038/s41567-020-0893-x} {\bibfield  {journal} {\bibinfo  {journal} {Nat. Phys.}\ }\textbf {\bibinfo {volume} {16}},\ \bibinfo {pages} {827} (\bibinfo {year} {2020})}\BibitemShut
  {NoStop}%
\bibitem [{\citenamefont {Ni}\ \emph {et~al.}(2023)\citenamefont {Ni}, \citenamefont {Li}, \citenamefont {Deng}, \citenamefont {Cai}, \citenamefont {Zhang}, \citenamefont {Wang}, \citenamefont {Yang}, \citenamefont {Yu}, \citenamefont {Yan}, \citenamefont {Liu}, \citenamefont {Zou}, \citenamefont {Sun}, \citenamefont {Zheng}, \citenamefont {Xu},\ and\ \citenamefont {Yu}}]{Ni2023}%
  \BibitemOpen
  \bibfield  {author} {\bibinfo {author} {\bibfnamefont {Z.}~\bibnamefont {Ni}}, \bibinfo {author} {\bibfnamefont {S.}~\bibnamefont {Li}}, \bibinfo {author} {\bibfnamefont {X.}~\bibnamefont {Deng}}, \bibinfo {author} {\bibfnamefont {Y.}~\bibnamefont {Cai}}, \bibinfo {author} {\bibfnamefont {L.}~\bibnamefont {Zhang}}, \bibinfo {author} {\bibfnamefont {W.}~\bibnamefont {Wang}}, \bibinfo {author} {\bibfnamefont {Z.-B.}\ \bibnamefont {Yang}}, \bibinfo {author} {\bibfnamefont {H.}~\bibnamefont {Yu}}, \bibinfo {author} {\bibfnamefont {F.}~\bibnamefont {Yan}}, \bibinfo {author} {\bibfnamefont {S.}~\bibnamefont {Liu}}, \bibinfo {author} {\bibfnamefont {C.-L.}\ \bibnamefont {Zou}}, \bibinfo {author} {\bibfnamefont {L.}~\bibnamefont {Sun}}, \bibinfo {author} {\bibfnamefont {S.-B.}\ \bibnamefont {Zheng}}, \bibinfo {author} {\bibfnamefont {Y.}~\bibnamefont {Xu}}, \ and\ \bibinfo {author} {\bibfnamefont {D.}~\bibnamefont {Yu}},\ }\bibfield  {title} {\enquote {\bibinfo {title} {Beating the break-even point with a
  discrete-variable-encoded logical qubit},}\ }\href {\doibase 10.1038/s41586-023-05784-4} {\bibfield  {journal} {\bibinfo  {journal} {Nature}\ }\textbf {\bibinfo {volume} {616}},\ \bibinfo {pages} {56} (\bibinfo {year} {2023})}\BibitemShut {NoStop}%
\bibitem [{\citenamefont {Mirrahimi}\ \emph {et~al.}(2014)\citenamefont {Mirrahimi}, \citenamefont {Leghtas}, \citenamefont {Albert}, \citenamefont {Touzard}, \citenamefont {Schoelkopf}, \citenamefont {Jiang},\ and\ \citenamefont {Devoret}}]{Mirrahimi2014}%
  \BibitemOpen
  \bibfield  {author} {\bibinfo {author} {\bibfnamefont {M.}~\bibnamefont {Mirrahimi}}, \bibinfo {author} {\bibfnamefont {Z.}~\bibnamefont {Leghtas}}, \bibinfo {author} {\bibfnamefont {V.~V.}\ \bibnamefont {Albert}}, \bibinfo {author} {\bibfnamefont {S.}~\bibnamefont {Touzard}}, \bibinfo {author} {\bibfnamefont {R.~J.}\ \bibnamefont {Schoelkopf}}, \bibinfo {author} {\bibfnamefont {L.}~\bibnamefont {Jiang}}, \ and\ \bibinfo {author} {\bibfnamefont {M.~H.}\ \bibnamefont {Devoret}},\ }\bibfield  {title} {\enquote {\bibinfo {title} {Dynamically protected cat-qubits: a new paradigm for universal quantum computation},}\ }\href {\doibase 10.1088/1367-2630/16/4/045014} {\bibfield  {journal} {\bibinfo  {journal} {New J. Phys.}\ }\textbf {\bibinfo {volume} {16}},\ \bibinfo {pages} {045014} (\bibinfo {year} {2014})}\BibitemShut {NoStop}%
\bibitem [{\citenamefont {Ofek}\ \emph {et~al.}(2016)\citenamefont {Ofek}, \citenamefont {Petrenko}, \citenamefont {Heeres}, \citenamefont {Reinhold}, \citenamefont {Leghtas}, \citenamefont {Vlastakis}, \citenamefont {Liu}, \citenamefont {Frunzio}, \citenamefont {Girvin}, \citenamefont {Jiang}, \citenamefont {Mirrahimi}, \citenamefont {Devoret},\ and\ \citenamefont {Schoelkopf}}]{Ofek2016}%
  \BibitemOpen
  \bibfield  {author} {\bibinfo {author} {\bibfnamefont {N.}~\bibnamefont {Ofek}}, \bibinfo {author} {\bibfnamefont {A.}~\bibnamefont {Petrenko}}, \bibinfo {author} {\bibfnamefont {R.}~\bibnamefont {Heeres}}, \bibinfo {author} {\bibfnamefont {P.}~\bibnamefont {Reinhold}}, \bibinfo {author} {\bibfnamefont {Z.}~\bibnamefont {Leghtas}}, \bibinfo {author} {\bibfnamefont {B.}~\bibnamefont {Vlastakis}}, \bibinfo {author} {\bibfnamefont {Y.}~\bibnamefont {Liu}}, \bibinfo {author} {\bibfnamefont {L.}~\bibnamefont {Frunzio}}, \bibinfo {author} {\bibfnamefont {S.~M.}\ \bibnamefont {Girvin}}, \bibinfo {author} {\bibfnamefont {L.}~\bibnamefont {Jiang}}, \bibinfo {author} {\bibfnamefont {M.}~\bibnamefont {Mirrahimi}}, \bibinfo {author} {\bibfnamefont {M.~H.}\ \bibnamefont {Devoret}}, \ and\ \bibinfo {author} {\bibfnamefont {R.~J.}\ \bibnamefont {Schoelkopf}},\ }\bibfield  {title} {\enquote {\bibinfo {title} {Extending the lifetime of a quantum bit with error correction in superconducting circuits},}\ }\href {\doibase
  10.1038/nature18949} {\bibfield  {journal} {\bibinfo  {journal} {Nature}\ }\textbf {\bibinfo {volume} {536}},\ \bibinfo {pages} {441} (\bibinfo {year} {2016})}\BibitemShut {NoStop}%
\bibitem [{\citenamefont {Gottesman}\ \emph {et~al.}(2001)\citenamefont {Gottesman}, \citenamefont {Kitaev},\ and\ \citenamefont {Preskill}}]{Gottesman2001}%
  \BibitemOpen
  \bibfield  {author} {\bibinfo {author} {\bibfnamefont {D.}~\bibnamefont {Gottesman}}, \bibinfo {author} {\bibfnamefont {A.}~\bibnamefont {Kitaev}}, \ and\ \bibinfo {author} {\bibfnamefont {J.}~\bibnamefont {Preskill}},\ }\bibfield  {title} {\enquote {\bibinfo {title} {Encoding a qubit in an oscillator},}\ }\href {\doibase 10.1103/PhysRevA.64.012310} {\bibfield  {journal} {\bibinfo  {journal} {Phys. Rev. A}\ }\textbf {\bibinfo {volume} {64}},\ \bibinfo {pages} {012310} (\bibinfo {year} {2001})}\BibitemShut {NoStop}%
\bibitem [{\citenamefont {Fl{\"u}hmann}\ \emph {et~al.}(2019)\citenamefont {Fl{\"u}hmann}, \citenamefont {Nguyen}, \citenamefont {Marinelli}, \citenamefont {Negnevitsky}, \citenamefont {Mehta},\ and\ \citenamefont {Home}}]{Fluhmann2019}%
  \BibitemOpen
  \bibfield  {author} {\bibinfo {author} {\bibfnamefont {C.}~\bibnamefont {Fl{\"u}hmann}}, \bibinfo {author} {\bibfnamefont {T.~L.}\ \bibnamefont {Nguyen}}, \bibinfo {author} {\bibfnamefont {M.}~\bibnamefont {Marinelli}}, \bibinfo {author} {\bibfnamefont {V.}~\bibnamefont {Negnevitsky}}, \bibinfo {author} {\bibfnamefont {K.}~\bibnamefont {Mehta}}, \ and\ \bibinfo {author} {\bibfnamefont {J.~P.}\ \bibnamefont {Home}},\ }\bibfield  {title} {\enquote {\bibinfo {title} {Encoding a qubit in a trapped-ion mechanical oscillator},}\ }\href {\doibase 10.1038/s41586-019-0960-6} {\bibfield  {journal} {\bibinfo  {journal} {Nature}\ }\textbf {\bibinfo {volume} {566}},\ \bibinfo {pages} {513} (\bibinfo {year} {2019})}\BibitemShut {NoStop}%
\bibitem [{\citenamefont {Campagne-Ibarcq}\ \emph {et~al.}(2020)\citenamefont {Campagne-Ibarcq}, \citenamefont {Eickbusch}, \citenamefont {Touzard}, \citenamefont {Zalys-Geller}, \citenamefont {Frattini}, \citenamefont {Sivak}, \citenamefont {Reinhold}, \citenamefont {Puri}, \citenamefont {Shankar}, \citenamefont {Schoelkopf}, \citenamefont {Frunzio}, \citenamefont {Mirrahimi},\ and\ \citenamefont {Devoret}}]{CampagneIbarcq2020}%
  \BibitemOpen
  \bibfield  {author} {\bibinfo {author} {\bibfnamefont {P.}~\bibnamefont {Campagne-Ibarcq}}, \bibinfo {author} {\bibfnamefont {A.}~\bibnamefont {Eickbusch}}, \bibinfo {author} {\bibfnamefont {S.}~\bibnamefont {Touzard}}, \bibinfo {author} {\bibfnamefont {E.}~\bibnamefont {Zalys-Geller}}, \bibinfo {author} {\bibfnamefont {N.~E.}\ \bibnamefont {Frattini}}, \bibinfo {author} {\bibfnamefont {V.~V.}\ \bibnamefont {Sivak}}, \bibinfo {author} {\bibfnamefont {P.}~\bibnamefont {Reinhold}}, \bibinfo {author} {\bibfnamefont {S.}~\bibnamefont {Puri}}, \bibinfo {author} {\bibfnamefont {S.}~\bibnamefont {Shankar}}, \bibinfo {author} {\bibfnamefont {R.~J.}\ \bibnamefont {Schoelkopf}}, \bibinfo {author} {\bibfnamefont {L.}~\bibnamefont {Frunzio}}, \bibinfo {author} {\bibfnamefont {M.}~\bibnamefont {Mirrahimi}}, \ and\ \bibinfo {author} {\bibfnamefont {M.~H.}\ \bibnamefont {Devoret}},\ }\bibfield  {title} {\enquote {\bibinfo {title} {Quantum error correction of a qubit encoded in grid states of an oscillator},}\ }\href
  {\doibase 10.1038/s41586-020-2603-3} {\bibfield  {journal} {\bibinfo  {journal} {Nature}\ }\textbf {\bibinfo {volume} {584}},\ \bibinfo {pages} {368} (\bibinfo {year} {2020})}\BibitemShut {NoStop}%
\bibitem [{\citenamefont {Sivak}\ \emph {et~al.}(2023)\citenamefont {Sivak}, \citenamefont {Eickbusch}, \citenamefont {Royer}, \citenamefont {Singh}, \citenamefont {Tsioutsios}, \citenamefont {Ganjam}, \citenamefont {Miano}, \citenamefont {Brock}, \citenamefont {Ding}, \citenamefont {Frunzio}, \citenamefont {Girvin}, \citenamefont {Schoelkopf},\ and\ \citenamefont {Devoret}}]{Sivak2023}%
  \BibitemOpen
  \bibfield  {author} {\bibinfo {author} {\bibfnamefont {V.~V.}\ \bibnamefont {Sivak}}, \bibinfo {author} {\bibfnamefont {A.}~\bibnamefont {Eickbusch}}, \bibinfo {author} {\bibfnamefont {B.}~\bibnamefont {Royer}}, \bibinfo {author} {\bibfnamefont {S.}~\bibnamefont {Singh}}, \bibinfo {author} {\bibfnamefont {I.}~\bibnamefont {Tsioutsios}}, \bibinfo {author} {\bibfnamefont {S.}~\bibnamefont {Ganjam}}, \bibinfo {author} {\bibfnamefont {A.}~\bibnamefont {Miano}}, \bibinfo {author} {\bibfnamefont {B.~L.}\ \bibnamefont {Brock}}, \bibinfo {author} {\bibfnamefont {A.~Z.}\ \bibnamefont {Ding}}, \bibinfo {author} {\bibfnamefont {L.}~\bibnamefont {Frunzio}}, \bibinfo {author} {\bibfnamefont {S.~M.}\ \bibnamefont {Girvin}}, \bibinfo {author} {\bibfnamefont {R.~J.}\ \bibnamefont {Schoelkopf}}, \ and\ \bibinfo {author} {\bibfnamefont {M.~H.}\ \bibnamefont {Devoret}},\ }\bibfield  {title} {\enquote {\bibinfo {title} {Real-time quantum error correction beyond break-even},}\ }\href {\doibase 10.1038/s41586-023-05782-6}
  {\bibfield  {journal} {\bibinfo  {journal} {Nature}\ }\textbf {\bibinfo {volume} {616}},\ \bibinfo {pages} {50} (\bibinfo {year} {2023})}\BibitemShut {NoStop}%
\bibitem [{\citenamefont {Sun}\ \emph {et~al.}()\citenamefont {Sun}, \citenamefont {Xu}, \citenamefont {Zhou}, \citenamefont {Hua}, \citenamefont {Wang}, \citenamefont {Zhou}, \citenamefont {Chen}, \citenamefont {de~Paula}, \citenamefont {Jie}, \citenamefont {Xue}, \citenamefont {Yu}, \citenamefont {Cai}, \citenamefont {Zou},\ and\ \citenamefont {Sun}}]{Sun2025}%
  \BibitemOpen
  \bibfield  {author} {\bibinfo {author} {\bibfnamefont {L.}~\bibnamefont {Sun}}, \bibinfo {author} {\bibfnamefont {Y.}~\bibnamefont {Xu}}, \bibinfo {author} {\bibfnamefont {Y.}~\bibnamefont {Zhou}}, \bibinfo {author} {\bibfnamefont {Z.}~\bibnamefont {Hua}}, \bibinfo {author} {\bibfnamefont {W.}~\bibnamefont {Wang}}, \bibinfo {author} {\bibfnamefont {J.}~\bibnamefont {Zhou}}, \bibinfo {author} {\bibfnamefont {Z.-J.}\ \bibnamefont {Chen}}, \bibinfo {author} {\bibfnamefont {L.~Z.}\ \bibnamefont {de~Paula}}, \bibinfo {author} {\bibfnamefont {Q.-X.}\ \bibnamefont {Jie}}, \bibinfo {author} {\bibfnamefont {G.}~\bibnamefont {Xue}}, \bibinfo {author} {\bibfnamefont {H.}~\bibnamefont {Yu}}, \bibinfo {author} {\bibfnamefont {W.}~\bibnamefont {Cai}}, \bibinfo {author} {\bibfnamefont {C.-L.}\ \bibnamefont {Zou}}, \ and\ \bibinfo {author} {\bibfnamefont {L.}~\bibnamefont {Sun}},\ }\bibfield  {title} {\enquote {\bibinfo {title} {Extending coherence time beyond break-even point using only drives and dissipation},}\
  }\href@noop {} {\ }\bibinfo {note} {\href{https://arxiv.org/abs/2509.22191}{arXiv:2509.22191} (2025)}\BibitemShut {NoStop}%
\bibitem [{\citenamefont {Ni}\ \emph {et~al.}()\citenamefont {Ni}, \citenamefont {Hu}, \citenamefont {Cai}, \citenamefont {Zhang}, \citenamefont {Mai}, \citenamefont {Deng}, \citenamefont {Zheng}, \citenamefont {Liu}, \citenamefont {Zheng}, \citenamefont {Xu},\ and\ \citenamefont {Yu}}]{Ni2025}%
  \BibitemOpen
  \bibfield  {author} {\bibinfo {author} {\bibfnamefont {Z.}~\bibnamefont {Ni}}, \bibinfo {author} {\bibfnamefont {L.}~\bibnamefont {Hu}}, \bibinfo {author} {\bibfnamefont {Y.}~\bibnamefont {Cai}}, \bibinfo {author} {\bibfnamefont {L.}~\bibnamefont {Zhang}}, \bibinfo {author} {\bibfnamefont {J.}~\bibnamefont {Mai}}, \bibinfo {author} {\bibfnamefont {X.}~\bibnamefont {Deng}}, \bibinfo {author} {\bibfnamefont {P.}~\bibnamefont {Zheng}}, \bibinfo {author} {\bibfnamefont {S.}~\bibnamefont {Liu}}, \bibinfo {author} {\bibfnamefont {S.-B.}\ \bibnamefont {Zheng}}, \bibinfo {author} {\bibfnamefont {Y.}~\bibnamefont {Xu}}, \ and\ \bibinfo {author} {\bibfnamefont {D.}~\bibnamefont {Yu}},\ }\bibfield  {title} {\enquote {\bibinfo {title} {Autonomous quantum error correction beyond break-even and its metrological application},}\ }\href@noop {} {\ }\bibinfo {note} {\href{https://arxiv.org/abs/2509.26042}{arXiv:2509.26042} (2025)}\BibitemShut {NoStop}%
\bibitem [{\citenamefont {Shirol}\ \emph {et~al.}(2026)\citenamefont {Shirol}, \citenamefont {van Geldern}, \citenamefont {Xi},\ and\ \citenamefont {Wang}}]{Shirol2026}%
  \BibitemOpen
  \bibfield  {author} {\bibinfo {author} {\bibfnamefont {S.}~\bibnamefont {Shirol}}, \bibinfo {author} {\bibfnamefont {S.}~\bibnamefont {van Geldern}}, \bibinfo {author} {\bibfnamefont {H.}~\bibnamefont {Xi}}, \ and\ \bibinfo {author} {\bibfnamefont {C.}~\bibnamefont {Wang}},\ }\bibfield  {title} {\enquote {\bibinfo {title} {Passive quantum error correction of photon loss at breakeven},}\ }\href {\doibase 10.1103/nvbm-97vs} {\bibfield  {journal} {\bibinfo  {journal} {Phys. Rev. X}\ }\textbf {\bibinfo {volume} {16}},\ \bibinfo {pages} {021042} (\bibinfo {year} {2026})}\BibitemShut {NoStop}%
\bibitem [{\citenamefont {Putterman}\ \emph {et~al.}(2025)\citenamefont {Putterman}, \citenamefont {Noh}, \citenamefont {Hann}, \citenamefont {MacCabe}, \citenamefont {Aghaeimeibodi}, \citenamefont {Patel}, \citenamefont {Lee}, \citenamefont {Jones}, \citenamefont {Moradinejad}, \citenamefont {Rodriguez}, \citenamefont {Mahuli}, \citenamefont {Rose}, \citenamefont {Owens}, \citenamefont {Levine}, \citenamefont {Rosenfeld} \emph {et~al.}}]{Putterman2025}%
  \BibitemOpen
  \bibfield  {author} {\bibinfo {author} {\bibfnamefont {H.}~\bibnamefont {Putterman}}, \bibinfo {author} {\bibfnamefont {K.}~\bibnamefont {Noh}}, \bibinfo {author} {\bibfnamefont {C.~T.}\ \bibnamefont {Hann}}, \bibinfo {author} {\bibfnamefont {G.~S.}\ \bibnamefont {MacCabe}}, \bibinfo {author} {\bibfnamefont {S.}~\bibnamefont {Aghaeimeibodi}}, \bibinfo {author} {\bibfnamefont {R.~N.}\ \bibnamefont {Patel}}, \bibinfo {author} {\bibfnamefont {M.}~\bibnamefont {Lee}}, \bibinfo {author} {\bibfnamefont {W.~M.}\ \bibnamefont {Jones}}, \bibinfo {author} {\bibfnamefont {H.}~\bibnamefont {Moradinejad}}, \bibinfo {author} {\bibfnamefont {R.}~\bibnamefont {Rodriguez}}, \bibinfo {author} {\bibfnamefont {N.}~\bibnamefont {Mahuli}}, \bibinfo {author} {\bibfnamefont {J.}~\bibnamefont {Rose}}, \bibinfo {author} {\bibfnamefont {J.~C.}\ \bibnamefont {Owens}}, \bibinfo {author} {\bibfnamefont {H.}~\bibnamefont {Levine}}, \bibinfo {author} {\bibfnamefont {E.}~\bibnamefont {Rosenfeld}},  \emph {et~al.},\ }\bibfield  {title}
  {\enquote {\bibinfo {title} {Hardware-efficient quantum error correction via concatenated bosonic qubits},}\ }\href {\doibase 10.1038/s41586-025-08642-7} {\bibfield  {journal} {\bibinfo  {journal} {Nature}\ }\textbf {\bibinfo {volume} {638}},\ \bibinfo {pages} {927} (\bibinfo {year} {2025})}\BibitemShut {NoStop}%
\bibitem [{\citenamefont {Noh}\ and\ \citenamefont {Chamberland}(2020)}]{Noh2020}%
  \BibitemOpen
  \bibfield  {author} {\bibinfo {author} {\bibfnamefont {K.}~\bibnamefont {Noh}}\ and\ \bibinfo {author} {\bibfnamefont {C.}~\bibnamefont {Chamberland}},\ }\bibfield  {title} {\enquote {\bibinfo {title} {Fault-tolerant bosonic quantum error correction with the surface--{Gottesman}--{Kitaev}--{Preskill} code},}\ }\href {\doibase 10.1103/PhysRevA.101.012316} {\bibfield  {journal} {\bibinfo  {journal} {Phys. Rev. A}\ }\textbf {\bibinfo {volume} {101}},\ \bibinfo {pages} {012316} (\bibinfo {year} {2020})}\BibitemShut {NoStop}%
\bibitem [{\citenamefont {Noh}\ \emph {et~al.}(2022)\citenamefont {Noh}, \citenamefont {Chamberland},\ and\ \citenamefont {Brand{\~a}o}}]{Noh2022}%
  \BibitemOpen
  \bibfield  {author} {\bibinfo {author} {\bibfnamefont {K.}~\bibnamefont {Noh}}, \bibinfo {author} {\bibfnamefont {C.}~\bibnamefont {Chamberland}}, \ and\ \bibinfo {author} {\bibfnamefont {F.~G. S.~L.}\ \bibnamefont {Brand{\~a}o}},\ }\bibfield  {title} {\enquote {\bibinfo {title} {Low-overhead fault-tolerant quantum error correction with the surface-{GKP} code},}\ }\href {\doibase 10.1103/PRXQuantum.3.010315} {\bibfield  {journal} {\bibinfo  {journal} {PRX Quantum}\ }\textbf {\bibinfo {volume} {3}},\ \bibinfo {pages} {010315} (\bibinfo {year} {2022})}\BibitemShut {NoStop}%
\bibitem [{\citenamefont {Albert}\ \emph {et~al.}(2018)\citenamefont {Albert}, \citenamefont {Noh}, \citenamefont {Duivenvoorden}, \citenamefont {Young}, \citenamefont {Brierley}, \citenamefont {Reinhold}, \citenamefont {Vuillot}, \citenamefont {Li}, \citenamefont {Shen}, \citenamefont {Girvin}, \citenamefont {Terhal},\ and\ \citenamefont {Jiang}}]{Albert2018}%
  \BibitemOpen
  \bibfield  {author} {\bibinfo {author} {\bibfnamefont {V.~V.}\ \bibnamefont {Albert}}, \bibinfo {author} {\bibfnamefont {K.}~\bibnamefont {Noh}}, \bibinfo {author} {\bibfnamefont {K.}~\bibnamefont {Duivenvoorden}}, \bibinfo {author} {\bibfnamefont {D.~J.}\ \bibnamefont {Young}}, \bibinfo {author} {\bibfnamefont {R.~T.}\ \bibnamefont {Brierley}}, \bibinfo {author} {\bibfnamefont {P.}~\bibnamefont {Reinhold}}, \bibinfo {author} {\bibfnamefont {C.}~\bibnamefont {Vuillot}}, \bibinfo {author} {\bibfnamefont {L.}~\bibnamefont {Li}}, \bibinfo {author} {\bibfnamefont {C.}~\bibnamefont {Shen}}, \bibinfo {author} {\bibfnamefont {S.~M.}\ \bibnamefont {Girvin}}, \bibinfo {author} {\bibfnamefont {B.~M.}\ \bibnamefont {Terhal}}, \ and\ \bibinfo {author} {\bibfnamefont {L.}~\bibnamefont {Jiang}},\ }\bibfield  {title} {\enquote {\bibinfo {title} {Performance and structure of single-mode bosonic codes},}\ }\href {\doibase 10.1103/PhysRevA.97.032346} {\bibfield  {journal} {\bibinfo  {journal} {Phys. Rev. A}\ }\textbf
  {\bibinfo {volume} {97}},\ \bibinfo {pages} {032346} (\bibinfo {year} {2018})}\BibitemShut {NoStop}%
\bibitem [{\citenamefont {Blais}\ \emph {et~al.}(2004)\citenamefont {Blais}, \citenamefont {Huang}, \citenamefont {Wallraff}, \citenamefont {Girvin},\ and\ \citenamefont {Schoelkopf}}]{Blais2004}%
  \BibitemOpen
  \bibfield  {author} {\bibinfo {author} {\bibfnamefont {A.}~\bibnamefont {Blais}}, \bibinfo {author} {\bibfnamefont {R.-S.}\ \bibnamefont {Huang}}, \bibinfo {author} {\bibfnamefont {A.}~\bibnamefont {Wallraff}}, \bibinfo {author} {\bibfnamefont {S.~M.}\ \bibnamefont {Girvin}}, \ and\ \bibinfo {author} {\bibfnamefont {R.~J.}\ \bibnamefont {Schoelkopf}},\ }\bibfield  {title} {\enquote {\bibinfo {title} {Cavity quantum electrodynamics for superconducting electrical circuits: An architecture for quantum computation},}\ }\href {\doibase 10.1103/PhysRevA.69.062320} {\bibfield  {journal} {\bibinfo  {journal} {Phys. Rev. A}\ }\textbf {\bibinfo {volume} {69}},\ \bibinfo {pages} {062320} (\bibinfo {year} {2004})}\BibitemShut {NoStop}%
\bibitem [{\citenamefont {Blais}\ \emph {et~al.}(2021)\citenamefont {Blais}, \citenamefont {Grimsmo}, \citenamefont {Girvin},\ and\ \citenamefont {Wallraff}}]{Blais2021}%
  \BibitemOpen
  \bibfield  {author} {\bibinfo {author} {\bibfnamefont {A.}~\bibnamefont {Blais}}, \bibinfo {author} {\bibfnamefont {A.~L.}\ \bibnamefont {Grimsmo}}, \bibinfo {author} {\bibfnamefont {S.~M.}\ \bibnamefont {Girvin}}, \ and\ \bibinfo {author} {\bibfnamefont {A.}~\bibnamefont {Wallraff}},\ }\bibfield  {title} {\enquote {\bibinfo {title} {Circuit quantum electrodynamics},}\ }\href {\doibase 10.1103/RevModPhys.93.025005} {\bibfield  {journal} {\bibinfo  {journal} {Rev. Mod. Phys.}\ }\textbf {\bibinfo {volume} {93}},\ \bibinfo {pages} {025005} (\bibinfo {year} {2021})}\BibitemShut {NoStop}%
\bibitem [{\citenamefont {Chapman}\ \emph {et~al.}(2023)\citenamefont {Chapman}, \citenamefont {de~Graaf}, \citenamefont {Xue}, \citenamefont {Teoh}, \citenamefont {Curtis}, \citenamefont {Tsunoda}, \citenamefont {Koottandavida}, \citenamefont {Frunzio}, \citenamefont {Girvin},\ and\ \citenamefont {Schoelkopf}}]{Chapman2023}%
  \BibitemOpen
  \bibfield  {author} {\bibinfo {author} {\bibfnamefont {B.~J.}\ \bibnamefont {Chapman}}, \bibinfo {author} {\bibfnamefont {S.~J.}\ \bibnamefont {de~Graaf}}, \bibinfo {author} {\bibfnamefont {S.~H.}\ \bibnamefont {Xue}}, \bibinfo {author} {\bibfnamefont {J.~D.}\ \bibnamefont {Teoh}}, \bibinfo {author} {\bibfnamefont {J.~C.}\ \bibnamefont {Curtis}}, \bibinfo {author} {\bibfnamefont {T.}~\bibnamefont {Tsunoda}}, \bibinfo {author} {\bibfnamefont {A.}~\bibnamefont {Koottandavida}}, \bibinfo {author} {\bibfnamefont {L.}~\bibnamefont {Frunzio}}, \bibinfo {author} {\bibfnamefont {S.~M.}\ \bibnamefont {Girvin}}, \ and\ \bibinfo {author} {\bibfnamefont {R.~J.}\ \bibnamefont {Schoelkopf}},\ }\bibfield  {title} {\enquote {\bibinfo {title} {High-on-off-ratio beam-splitter interaction for gates on bosonically encoded qubits},}\ }\href {\doibase 10.1103/PRXQuantum.4.020355} {\bibfield  {journal} {\bibinfo  {journal} {PRX Quantum}\ }\textbf {\bibinfo {volume} {4}},\ \bibinfo {pages} {020355} (\bibinfo {year}
  {2023})}\BibitemShut {NoStop}%
\bibitem [{\citenamefont {Kirchmair}\ \emph {et~al.}(2013)\citenamefont {Kirchmair}, \citenamefont {Vlastakis}, \citenamefont {Leghtas}, \citenamefont {Nigg}, \citenamefont {Paik}, \citenamefont {Ginossar}, \citenamefont {Mirrahimi}, \citenamefont {Frunzio}, \citenamefont {Girvin},\ and\ \citenamefont {Schoelkopf}}]{Kirchmair2013}%
  \BibitemOpen
  \bibfield  {author} {\bibinfo {author} {\bibfnamefont {G.}~\bibnamefont {Kirchmair}}, \bibinfo {author} {\bibfnamefont {B.}~\bibnamefont {Vlastakis}}, \bibinfo {author} {\bibfnamefont {Z.}~\bibnamefont {Leghtas}}, \bibinfo {author} {\bibfnamefont {S.~E.}\ \bibnamefont {Nigg}}, \bibinfo {author} {\bibfnamefont {H.}~\bibnamefont {Paik}}, \bibinfo {author} {\bibfnamefont {E.}~\bibnamefont {Ginossar}}, \bibinfo {author} {\bibfnamefont {M.}~\bibnamefont {Mirrahimi}}, \bibinfo {author} {\bibfnamefont {L.}~\bibnamefont {Frunzio}}, \bibinfo {author} {\bibfnamefont {S.~M.}\ \bibnamefont {Girvin}}, \ and\ \bibinfo {author} {\bibfnamefont {R.~J.}\ \bibnamefont {Schoelkopf}},\ }\bibfield  {title} {\enquote {\bibinfo {title} {Observation of quantum state collapse and revival due to the single-photon {Kerr} effect},}\ }\href {\doibase 10.1038/nature11902} {\bibfield  {journal} {\bibinfo  {journal} {Nature}\ }\textbf {\bibinfo {volume} {495}},\ \bibinfo {pages} {205} (\bibinfo {year} {2013})}\BibitemShut {NoStop}%
\bibitem [{\citenamefont {Grimm}\ \emph {et~al.}(2020)\citenamefont {Grimm}, \citenamefont {Frattini}, \citenamefont {Puri}, \citenamefont {Mundhada}, \citenamefont {Touzard}, \citenamefont {Mirrahimi}, \citenamefont {Girvin}, \citenamefont {Shankar},\ and\ \citenamefont {Devoret}}]{Grimm2020}%
  \BibitemOpen
  \bibfield  {author} {\bibinfo {author} {\bibfnamefont {A.}~\bibnamefont {Grimm}}, \bibinfo {author} {\bibfnamefont {N.~E.}\ \bibnamefont {Frattini}}, \bibinfo {author} {\bibfnamefont {S.}~\bibnamefont {Puri}}, \bibinfo {author} {\bibfnamefont {S.~O.}\ \bibnamefont {Mundhada}}, \bibinfo {author} {\bibfnamefont {S.}~\bibnamefont {Touzard}}, \bibinfo {author} {\bibfnamefont {M.}~\bibnamefont {Mirrahimi}}, \bibinfo {author} {\bibfnamefont {S.~M.}\ \bibnamefont {Girvin}}, \bibinfo {author} {\bibfnamefont {S.}~\bibnamefont {Shankar}}, \ and\ \bibinfo {author} {\bibfnamefont {M.~H.}\ \bibnamefont {Devoret}},\ }\bibfield  {title} {\enquote {\bibinfo {title} {Stabilization and operation of a {Kerr}-cat qubit},}\ }\href {\doibase 10.1038/s41586-020-2587-z} {\bibfield  {journal} {\bibinfo  {journal} {Nature}\ }\textbf {\bibinfo {volume} {584}},\ \bibinfo {pages} {205} (\bibinfo {year} {2020})}\BibitemShut {NoStop}%
\bibitem [{\citenamefont {Martinis}\ \emph {et~al.}(2005)\citenamefont {Martinis}, \citenamefont {Cooper}, \citenamefont {McDermott}, \citenamefont {Steffen}, \citenamefont {Ansmann}, \citenamefont {Osborn}, \citenamefont {Cicak}, \citenamefont {Oh}, \citenamefont {Pappas}, \citenamefont {Simmonds},\ and\ \citenamefont {Yu}}]{Martinis2005}%
  \BibitemOpen
  \bibfield  {author} {\bibinfo {author} {\bibfnamefont {J.~M.}\ \bibnamefont {Martinis}}, \bibinfo {author} {\bibfnamefont {K.~B.}\ \bibnamefont {Cooper}}, \bibinfo {author} {\bibfnamefont {R.}~\bibnamefont {McDermott}}, \bibinfo {author} {\bibfnamefont {M.}~\bibnamefont {Steffen}}, \bibinfo {author} {\bibfnamefont {M.}~\bibnamefont {Ansmann}}, \bibinfo {author} {\bibfnamefont {K.~D.}\ \bibnamefont {Osborn}}, \bibinfo {author} {\bibfnamefont {K.}~\bibnamefont {Cicak}}, \bibinfo {author} {\bibfnamefont {S.}~\bibnamefont {Oh}}, \bibinfo {author} {\bibfnamefont {D.~P.}\ \bibnamefont {Pappas}}, \bibinfo {author} {\bibfnamefont {R.~W.}\ \bibnamefont {Simmonds}}, \ and\ \bibinfo {author} {\bibfnamefont {C.~C.}\ \bibnamefont {Yu}},\ }\bibfield  {title} {\enquote {\bibinfo {title} {Decoherence in {Josephson} qubits from dielectric loss},}\ }\href {\doibase 10.1103/PhysRevLett.95.210503} {\bibfield  {journal} {\bibinfo  {journal} {Phys. Rev. Lett.}\ }\textbf {\bibinfo {volume} {95}},\ \bibinfo {pages} {210503}
  (\bibinfo {year} {2005})}\BibitemShut {NoStop}%
\bibitem [{\citenamefont {Simmonds}\ \emph {et~al.}(2004)\citenamefont {Simmonds}, \citenamefont {Lang}, \citenamefont {Hite}, \citenamefont {Nam}, \citenamefont {Pappas},\ and\ \citenamefont {Martinis}}]{Simmonds2004}%
  \BibitemOpen
  \bibfield  {author} {\bibinfo {author} {\bibfnamefont {R.~W.}\ \bibnamefont {Simmonds}}, \bibinfo {author} {\bibfnamefont {K.~M.}\ \bibnamefont {Lang}}, \bibinfo {author} {\bibfnamefont {D.~A.}\ \bibnamefont {Hite}}, \bibinfo {author} {\bibfnamefont {S.}~\bibnamefont {Nam}}, \bibinfo {author} {\bibfnamefont {D.~P.}\ \bibnamefont {Pappas}}, \ and\ \bibinfo {author} {\bibfnamefont {J.~M.}\ \bibnamefont {Martinis}},\ }\bibfield  {title} {\enquote {\bibinfo {title} {Decoherence in {Josephson} phase qubits from junction resonators},}\ }\href {\doibase 10.1103/PhysRevLett.93.077003} {\bibfield  {journal} {\bibinfo  {journal} {Phys. Rev. Lett.}\ }\textbf {\bibinfo {volume} {93}},\ \bibinfo {pages} {077003} (\bibinfo {year} {2004})}\BibitemShut {NoStop}%
\bibitem [{\citenamefont {Catelani}\ \emph {et~al.}(2011)\citenamefont {Catelani}, \citenamefont {Schoelkopf}, \citenamefont {Devoret},\ and\ \citenamefont {Glazman}}]{Catelani2011}%
  \BibitemOpen
  \bibfield  {author} {\bibinfo {author} {\bibfnamefont {G.}~\bibnamefont {Catelani}}, \bibinfo {author} {\bibfnamefont {R.~J.}\ \bibnamefont {Schoelkopf}}, \bibinfo {author} {\bibfnamefont {M.~H.}\ \bibnamefont {Devoret}}, \ and\ \bibinfo {author} {\bibfnamefont {L.~I.}\ \bibnamefont {Glazman}},\ }\bibfield  {title} {\enquote {\bibinfo {title} {Relaxation and frequency shifts induced by quasiparticles in superconducting qubits},}\ }\href {\doibase 10.1103/PhysRevB.84.064517} {\bibfield  {journal} {\bibinfo  {journal} {Phys. Rev. B}\ }\textbf {\bibinfo {volume} {84}},\ \bibinfo {pages} {064517} (\bibinfo {year} {2011})}\BibitemShut {NoStop}%
\bibitem [{\citenamefont {Clerk}\ \emph {et~al.}(2010)\citenamefont {Clerk}, \citenamefont {Devoret}, \citenamefont {Girvin}, \citenamefont {Marquardt},\ and\ \citenamefont {Schoelkopf}}]{Clerk2010}%
  \BibitemOpen
  \bibfield  {author} {\bibinfo {author} {\bibfnamefont {A.~A.}\ \bibnamefont {Clerk}}, \bibinfo {author} {\bibfnamefont {M.~H.}\ \bibnamefont {Devoret}}, \bibinfo {author} {\bibfnamefont {S.~M.}\ \bibnamefont {Girvin}}, \bibinfo {author} {\bibfnamefont {F.}~\bibnamefont {Marquardt}}, \ and\ \bibinfo {author} {\bibfnamefont {R.~J.}\ \bibnamefont {Schoelkopf}},\ }\bibfield  {title} {\enquote {\bibinfo {title} {Introduction to quantum noise, measurement, and amplification},}\ }\href {\doibase 10.1103/RevModPhys.82.1155} {\bibfield  {journal} {\bibinfo  {journal} {Rev. Mod. Phys.}\ }\textbf {\bibinfo {volume} {82}},\ \bibinfo {pages} {1155} (\bibinfo {year} {2010})}\BibitemShut {NoStop}%
\bibitem [{\citenamefont {Martinis}\ \emph {et~al.}(2009)\citenamefont {Martinis}, \citenamefont {Ansmann},\ and\ \citenamefont {Aumentado}}]{Martinis2009}%
  \BibitemOpen
  \bibfield  {author} {\bibinfo {author} {\bibfnamefont {J.~M.}\ \bibnamefont {Martinis}}, \bibinfo {author} {\bibfnamefont {M.}~\bibnamefont {Ansmann}}, \ and\ \bibinfo {author} {\bibfnamefont {J.}~\bibnamefont {Aumentado}},\ }\bibfield  {title} {\enquote {\bibinfo {title} {Energy decay in superconducting {Josephson}-junction qubits from nonequilibrium quasiparticle excitations},}\ }\href {\doibase 10.1103/PhysRevLett.103.097002} {\bibfield  {journal} {\bibinfo  {journal} {Phys. Rev. Lett.}\ }\textbf {\bibinfo {volume} {103}},\ \bibinfo {pages} {097002} (\bibinfo {year} {2009})}\BibitemShut {NoStop}%
\bibitem [{\citenamefont {Siddiqi}(2021)}]{Siddiqi2021}%
  \BibitemOpen
  \bibfield  {author} {\bibinfo {author} {\bibfnamefont {I.}~\bibnamefont {Siddiqi}},\ }\bibfield  {title} {\enquote {\bibinfo {title} {Engineering high-coherence superconducting qubits},}\ }\href {\doibase 10.1038/s41578-021-00370-4} {\bibfield  {journal} {\bibinfo  {journal} {Nat. Rev. Mater.}\ }\textbf {\bibinfo {volume} {6}},\ \bibinfo {pages} {875} (\bibinfo {year} {2021})}\BibitemShut {NoStop}%
\bibitem [{\citenamefont {Yan}\ \emph {et~al.}(2018)\citenamefont {Yan}, \citenamefont {Gustavsson}, \citenamefont {Kamal}, \citenamefont {Birenbaum}, \citenamefont {Sears}, \citenamefont {Hover}, \citenamefont {Gudmundsen}, \citenamefont {Rosenberg}, \citenamefont {Samach}, \citenamefont {Weber}, \citenamefont {Orlando}, \citenamefont {Clarke}, \citenamefont {Kerman},\ and\ \citenamefont {Oliver}}]{Yan2018}%
  \BibitemOpen
  \bibfield  {author} {\bibinfo {author} {\bibfnamefont {F.}~\bibnamefont {Yan}}, \bibinfo {author} {\bibfnamefont {S.}~\bibnamefont {Gustavsson}}, \bibinfo {author} {\bibfnamefont {A.}~\bibnamefont {Kamal}}, \bibinfo {author} {\bibfnamefont {J.}~\bibnamefont {Birenbaum}}, \bibinfo {author} {\bibfnamefont {A.~P.}\ \bibnamefont {Sears}}, \bibinfo {author} {\bibfnamefont {D.}~\bibnamefont {Hover}}, \bibinfo {author} {\bibfnamefont {T.~J.}\ \bibnamefont {Gudmundsen}}, \bibinfo {author} {\bibfnamefont {D.}~\bibnamefont {Rosenberg}}, \bibinfo {author} {\bibfnamefont {G.}~\bibnamefont {Samach}}, \bibinfo {author} {\bibfnamefont {S.}~\bibnamefont {Weber}}, \bibinfo {author} {\bibfnamefont {T.~P.}\ \bibnamefont {Orlando}}, \bibinfo {author} {\bibfnamefont {J.}~\bibnamefont {Clarke}}, \bibinfo {author} {\bibfnamefont {A.~J.}\ \bibnamefont {Kerman}}, \ and\ \bibinfo {author} {\bibfnamefont {W.~D.}\ \bibnamefont {Oliver}},\ }\bibfield  {title} {\enquote {\bibinfo {title} {Distinguishing coherent and thermal photon noise
  in a circuit quantum electrodynamical system},}\ }\href {\doibase 10.1103/PhysRevLett.120.260504} {\bibfield  {journal} {\bibinfo  {journal} {Phys. Rev. Lett.}\ }\textbf {\bibinfo {volume} {120}},\ \bibinfo {pages} {260504} (\bibinfo {year} {2018})}\BibitemShut {NoStop}%
\bibitem [{\citenamefont {Breuer}\ and\ \citenamefont {Petruccione}(2002)}]{Breuer2002}%
  \BibitemOpen
  \bibfield  {author} {\bibinfo {author} {\bibfnamefont {H.-P.}\ \bibnamefont {Breuer}}\ and\ \bibinfo {author} {\bibfnamefont {F.}~\bibnamefont {Petruccione}},\ }\href@noop {} {\emph {\bibinfo {title} {The Theory of Open Quantum Systems}}}\ (\bibinfo  {publisher} {Oxford University Press},\ \bibinfo {address} {Oxford},\ \bibinfo {year} {2002})\BibitemShut {NoStop}%
\bibitem [{\citenamefont {Breuer}\ \emph {et~al.}(2016)\citenamefont {Breuer}, \citenamefont {Laine}, \citenamefont {Piilo},\ and\ \citenamefont {Vacchini}}]{Breuer2016}%
  \BibitemOpen
  \bibfield  {author} {\bibinfo {author} {\bibfnamefont {H.-P.}\ \bibnamefont {Breuer}}, \bibinfo {author} {\bibfnamefont {E.-M.}\ \bibnamefont {Laine}}, \bibinfo {author} {\bibfnamefont {J.}~\bibnamefont {Piilo}}, \ and\ \bibinfo {author} {\bibfnamefont {B.}~\bibnamefont {Vacchini}},\ }\bibfield  {title} {\enquote {\bibinfo {title} {Colloquium: Non-{Markovian} dynamics in open quantum systems},}\ }\href {\doibase 10.1103/RevModPhys.88.021002} {\bibfield  {journal} {\bibinfo  {journal} {Rev. Mod. Phys.}\ }\textbf {\bibinfo {volume} {88}},\ \bibinfo {pages} {021002} (\bibinfo {year} {2016})}\BibitemShut {NoStop}%
\bibitem [{\citenamefont {Brillant}\ \emph {et~al.}(2025)\citenamefont {Brillant}, \citenamefont {Groszkowski}, \citenamefont {Seif}, \citenamefont {Koch},\ and\ \citenamefont {Clerk}}]{Brillant2025}%
  \BibitemOpen
  \bibfield  {author} {\bibinfo {author} {\bibfnamefont {A.}~\bibnamefont {Brillant}}, \bibinfo {author} {\bibfnamefont {P.}~\bibnamefont {Groszkowski}}, \bibinfo {author} {\bibfnamefont {A.}~\bibnamefont {Seif}}, \bibinfo {author} {\bibfnamefont {J.}~\bibnamefont {Koch}}, \ and\ \bibinfo {author} {\bibfnamefont {A.~A.}\ \bibnamefont {Clerk}},\ }\bibfield  {title} {\enquote {\bibinfo {title} {Randomized benchmarking with non-{Markovian} noise and realistic finite-time gates},}\ }\href {\doibase 10.1103/tmfd-qc5q} {\bibfield  {journal} {\bibinfo  {journal} {Phys. Rev. Lett.}\ }\textbf {\bibinfo {volume} {135}},\ \bibinfo {pages} {070601} (\bibinfo {year} {2025})}\BibitemShut {NoStop}%
\bibitem [{\citenamefont {Paladino}\ \emph {et~al.}(2014)\citenamefont {Paladino}, \citenamefont {Galperin}, \citenamefont {Falci},\ and\ \citenamefont {Altshuler}}]{Paladino2014}%
  \BibitemOpen
  \bibfield  {author} {\bibinfo {author} {\bibfnamefont {E.}~\bibnamefont {Paladino}}, \bibinfo {author} {\bibfnamefont {Y.~M.}\ \bibnamefont {Galperin}}, \bibinfo {author} {\bibfnamefont {G.}~\bibnamefont {Falci}}, \ and\ \bibinfo {author} {\bibfnamefont {B.~L.}\ \bibnamefont {Altshuler}},\ }\bibfield  {title} {\enquote {\bibinfo {title} {{$1/f$} noise: Implications for solid-state quantum information},}\ }\href {\doibase 10.1103/RevModPhys.86.361} {\bibfield  {journal} {\bibinfo  {journal} {Rev. Mod. Phys.}\ }\textbf {\bibinfo {volume} {86}},\ \bibinfo {pages} {361} (\bibinfo {year} {2014})}\BibitemShut {NoStop}%
\bibitem [{\citenamefont {Figueroa-Romero}\ \emph {et~al.}(2021)\citenamefont {Figueroa-Romero}, \citenamefont {Modi}, \citenamefont {Harris}, \citenamefont {Stace},\ and\ \citenamefont {Hsieh}}]{Figueroa2021}%
  \BibitemOpen
  \bibfield  {author} {\bibinfo {author} {\bibfnamefont {P.}~\bibnamefont {Figueroa-Romero}}, \bibinfo {author} {\bibfnamefont {K.}~\bibnamefont {Modi}}, \bibinfo {author} {\bibfnamefont {R.~J.}\ \bibnamefont {Harris}}, \bibinfo {author} {\bibfnamefont {T.~M.}\ \bibnamefont {Stace}}, \ and\ \bibinfo {author} {\bibfnamefont {M.-H.}\ \bibnamefont {Hsieh}},\ }\bibfield  {title} {\enquote {\bibinfo {title} {Randomized benchmarking for non-{Markovian} noise},}\ }\href {\doibase 10.1103/PRXQuantum.2.040351} {\bibfield  {journal} {\bibinfo  {journal} {PRX Quantum}\ }\textbf {\bibinfo {volume} {2}},\ \bibinfo {pages} {040351} (\bibinfo {year} {2021})}\BibitemShut {NoStop}%
\bibitem [{\citenamefont {Reagor}\ \emph {et~al.}(2016)\citenamefont {Reagor}, \citenamefont {Pfaff}, \citenamefont {Axline}, \citenamefont {Heeres}, \citenamefont {Ofek}, \citenamefont {Sliwa}, \citenamefont {Holland}, \citenamefont {Wang}, \citenamefont {Blumoff}, \citenamefont {Chou}, \citenamefont {Hatridge}, \citenamefont {Frunzio}, \citenamefont {Devoret}, \citenamefont {Jiang},\ and\ \citenamefont {Schoelkopf}}]{Reagor2016}%
  \BibitemOpen
  \bibfield  {author} {\bibinfo {author} {\bibfnamefont {M.}~\bibnamefont {Reagor}}, \bibinfo {author} {\bibfnamefont {W.}~\bibnamefont {Pfaff}}, \bibinfo {author} {\bibfnamefont {C.}~\bibnamefont {Axline}}, \bibinfo {author} {\bibfnamefont {R.~W.}\ \bibnamefont {Heeres}}, \bibinfo {author} {\bibfnamefont {N.}~\bibnamefont {Ofek}}, \bibinfo {author} {\bibfnamefont {K.}~\bibnamefont {Sliwa}}, \bibinfo {author} {\bibfnamefont {E.}~\bibnamefont {Holland}}, \bibinfo {author} {\bibfnamefont {C.}~\bibnamefont {Wang}}, \bibinfo {author} {\bibfnamefont {J.}~\bibnamefont {Blumoff}}, \bibinfo {author} {\bibfnamefont {K.}~\bibnamefont {Chou}}, \bibinfo {author} {\bibfnamefont {M.~J.}\ \bibnamefont {Hatridge}}, \bibinfo {author} {\bibfnamefont {L.}~\bibnamefont {Frunzio}}, \bibinfo {author} {\bibfnamefont {M.~H.}\ \bibnamefont {Devoret}}, \bibinfo {author} {\bibfnamefont {L.}~\bibnamefont {Jiang}}, \ and\ \bibinfo {author} {\bibfnamefont {R.~J.}\ \bibnamefont {Schoelkopf}},\ }\bibfield  {title} {\enquote {\bibinfo
  {title} {Quantum memory with millisecond coherence in circuit {QED}},}\ }\href {\doibase 10.1103/PhysRevB.94.014506} {\bibfield  {journal} {\bibinfo  {journal} {Phys. Rev. B}\ }\textbf {\bibinfo {volume} {94}},\ \bibinfo {pages} {014506} (\bibinfo {year} {2016})}\BibitemShut {NoStop}%
\bibitem [{\citenamefont {Axline}\ \emph {et~al.}(2016)\citenamefont {Axline}, \citenamefont {Reagor}, \citenamefont {Heeres}, \citenamefont {Reinhold}, \citenamefont {Wang}, \citenamefont {Shain}, \citenamefont {Pfaff}, \citenamefont {Chu}, \citenamefont {Frunzio},\ and\ \citenamefont {Schoelkopf}}]{Axline2016}%
  \BibitemOpen
  \bibfield  {author} {\bibinfo {author} {\bibfnamefont {C.}~\bibnamefont {Axline}}, \bibinfo {author} {\bibfnamefont {M.}~\bibnamefont {Reagor}}, \bibinfo {author} {\bibfnamefont {R.}~\bibnamefont {Heeres}}, \bibinfo {author} {\bibfnamefont {P.}~\bibnamefont {Reinhold}}, \bibinfo {author} {\bibfnamefont {C.}~\bibnamefont {Wang}}, \bibinfo {author} {\bibfnamefont {K.}~\bibnamefont {Shain}}, \bibinfo {author} {\bibfnamefont {W.}~\bibnamefont {Pfaff}}, \bibinfo {author} {\bibfnamefont {Y.}~\bibnamefont {Chu}}, \bibinfo {author} {\bibfnamefont {L.}~\bibnamefont {Frunzio}}, \ and\ \bibinfo {author} {\bibfnamefont {R.~J.}\ \bibnamefont {Schoelkopf}},\ }\bibfield  {title} {\enquote {\bibinfo {title} {An architecture for integrating planar and {3D} {cQED} devices},}\ }\href {\doibase 10.1063/1.4959241} {\bibfield  {journal} {\bibinfo  {journal} {Appl. Phys. Lett.}\ }\textbf {\bibinfo {volume} {109}},\ \bibinfo {pages} {042601} (\bibinfo {year} {2016})}\BibitemShut {NoStop}%
\bibitem [{\citenamefont {Wang}\ \emph {et~al.}()\citenamefont {Wang}, \citenamefont {Ding}, \citenamefont {Wang}, \citenamefont {Zhang}, \citenamefont {Wang}, \citenamefont {Nuerbolati}, \citenamefont {Yang}, \citenamefont {Liang}, \citenamefont {Sun}, \citenamefont {Yu},\ and\ \citenamefont {Yan}}]{Wang2026}%
  \BibitemOpen
  \bibfield  {author} {\bibinfo {author} {\bibfnamefont {R.}~\bibnamefont {Wang}}, \bibinfo {author} {\bibfnamefont {J.}~\bibnamefont {Ding}}, \bibinfo {author} {\bibfnamefont {C.}~\bibnamefont {Wang}}, \bibinfo {author} {\bibfnamefont {Y.}~\bibnamefont {Zhang}}, \bibinfo {author} {\bibfnamefont {H.}~\bibnamefont {Wang}}, \bibinfo {author} {\bibfnamefont {W.}~\bibnamefont {Nuerbolati}}, \bibinfo {author} {\bibfnamefont {Z.}~\bibnamefont {Yang}}, \bibinfo {author} {\bibfnamefont {X.}~\bibnamefont {Liang}}, \bibinfo {author} {\bibfnamefont {W.}~\bibnamefont {Sun}}, \bibinfo {author} {\bibfnamefont {H.}~\bibnamefont {Yu}}, \ and\ \bibinfo {author} {\bibfnamefont {F.}~\bibnamefont {Yan}},\ }\bibfield  {title} {\enquote {\bibinfo {title} {Dressed-state relaxation in coupled qubits as a source of two-qubit gate errors},}\ }\href@noop {} {\ }\bibinfo {note} {\href{https://arxiv.org/abs/2601.11316}{arXiv:2601.11316} (2026)}\BibitemShut {NoStop}%
\bibitem [{\citenamefont {Wang}\ \emph {et~al.}(2008)\citenamefont {Wang}, \citenamefont {Hofheinz}, \citenamefont {Ansmann}, \citenamefont {Bialczak}, \citenamefont {Lucero}, \citenamefont {Neeley}, \citenamefont {O'Connell}, \citenamefont {Sank}, \citenamefont {Wenner}, \citenamefont {Cleland},\ and\ \citenamefont {Martinis}}]{Wang2008}%
  \BibitemOpen
  \bibfield  {author} {\bibinfo {author} {\bibfnamefont {H.}~\bibnamefont {Wang}}, \bibinfo {author} {\bibfnamefont {M.}~\bibnamefont {Hofheinz}}, \bibinfo {author} {\bibfnamefont {M.}~\bibnamefont {Ansmann}}, \bibinfo {author} {\bibfnamefont {R.~C.}\ \bibnamefont {Bialczak}}, \bibinfo {author} {\bibfnamefont {E.}~\bibnamefont {Lucero}}, \bibinfo {author} {\bibfnamefont {M.}~\bibnamefont {Neeley}}, \bibinfo {author} {\bibfnamefont {A.~D.}\ \bibnamefont {O'Connell}}, \bibinfo {author} {\bibfnamefont {D.}~\bibnamefont {Sank}}, \bibinfo {author} {\bibfnamefont {J.}~\bibnamefont {Wenner}}, \bibinfo {author} {\bibfnamefont {A.~N.}\ \bibnamefont {Cleland}}, \ and\ \bibinfo {author} {\bibfnamefont {J.~M.}\ \bibnamefont {Martinis}},\ }\bibfield  {title} {\enquote {\bibinfo {title} {Measurement of the decay of {Fock} states in a superconducting quantum circuit},}\ }\href {\doibase 10.1103/PhysRevLett.101.240401} {\bibfield  {journal} {\bibinfo  {journal} {Phys. Rev. Lett.}\ }\textbf {\bibinfo {volume} {101}},\
  \bibinfo {pages} {240401} (\bibinfo {year} {2008})}\BibitemShut {NoStop}%
\bibitem [{\citenamefont {Peterer}\ \emph {et~al.}(2015)\citenamefont {Peterer}, \citenamefont {Bader}, \citenamefont {Jin}, \citenamefont {Yan}, \citenamefont {Kamal}, \citenamefont {Gudmundsen}, \citenamefont {Leek}, \citenamefont {Orlando}, \citenamefont {Oliver},\ and\ \citenamefont {Gustavsson}}]{Peterer2015}%
  \BibitemOpen
  \bibfield  {author} {\bibinfo {author} {\bibfnamefont {M.~J.}\ \bibnamefont {Peterer}}, \bibinfo {author} {\bibfnamefont {S.~J.}\ \bibnamefont {Bader}}, \bibinfo {author} {\bibfnamefont {X.}~\bibnamefont {Jin}}, \bibinfo {author} {\bibfnamefont {F.}~\bibnamefont {Yan}}, \bibinfo {author} {\bibfnamefont {A.}~\bibnamefont {Kamal}}, \bibinfo {author} {\bibfnamefont {T.~J.}\ \bibnamefont {Gudmundsen}}, \bibinfo {author} {\bibfnamefont {P.~J.}\ \bibnamefont {Leek}}, \bibinfo {author} {\bibfnamefont {T.~P.}\ \bibnamefont {Orlando}}, \bibinfo {author} {\bibfnamefont {W.~D.}\ \bibnamefont {Oliver}}, \ and\ \bibinfo {author} {\bibfnamefont {S.}~\bibnamefont {Gustavsson}},\ }\bibfield  {title} {\enquote {\bibinfo {title} {Coherence and decay of higher energy levels of a superconducting transmon qubit},}\ }\href {\doibase 10.1103/PhysRevLett.114.010501} {\bibfield  {journal} {\bibinfo  {journal} {Phys. Rev. Lett.}\ }\textbf {\bibinfo {volume} {114}},\ \bibinfo {pages} {010501} (\bibinfo {year} {2015})}\BibitemShut
  {NoStop}%
\bibitem [{\citenamefont {Wang}\ \emph {et~al.}(2025)\citenamefont {Wang}, \citenamefont {Parker}, \citenamefont {Champion},\ and\ \citenamefont {Blok}}]{Wang2025}%
  \BibitemOpen
  \bibfield  {author} {\bibinfo {author} {\bibfnamefont {Z.}~\bibnamefont {Wang}}, \bibinfo {author} {\bibfnamefont {R.~W.}\ \bibnamefont {Parker}}, \bibinfo {author} {\bibfnamefont {E.}~\bibnamefont {Champion}}, \ and\ \bibinfo {author} {\bibfnamefont {M.~S.}\ \bibnamefont {Blok}},\ }\bibfield  {title} {\enquote {\bibinfo {title} {Systematic study of high {$E_J/E_C$} transmon qudits up to {$d=12$}},}\ }\href {\doibase 10.1103/PhysRevApplied.23.034046} {\bibfield  {journal} {\bibinfo  {journal} {Phys. Rev. Applied}\ }\textbf {\bibinfo {volume} {23}},\ \bibinfo {pages} {034046} (\bibinfo {year} {2025})}\BibitemShut {NoStop}%
\bibitem [{\citenamefont {Kim}\ \emph {et~al.}(2025)\citenamefont {Kim}, \citenamefont {Roy}, \citenamefont {You}, \citenamefont {Li}, \citenamefont {Lamm}, \citenamefont {Pronitchev}, \citenamefont {Bal}, \citenamefont {Garattoni}, \citenamefont {Crisa}, \citenamefont {Bafia}, \citenamefont {Kurkcuoglu}, \citenamefont {Pilipenko}, \citenamefont {Heidler}, \citenamefont {Bornman}, \citenamefont {van Zanten} \emph {et~al.}}]{Kim2025}%
  \BibitemOpen
  \bibfield  {author} {\bibinfo {author} {\bibfnamefont {T.}~\bibnamefont {Kim}}, \bibinfo {author} {\bibfnamefont {T.}~\bibnamefont {Roy}}, \bibinfo {author} {\bibfnamefont {X.}~\bibnamefont {You}}, \bibinfo {author} {\bibfnamefont {A.~C.~Y.}\ \bibnamefont {Li}}, \bibinfo {author} {\bibfnamefont {H.}~\bibnamefont {Lamm}}, \bibinfo {author} {\bibfnamefont {O.}~\bibnamefont {Pronitchev}}, \bibinfo {author} {\bibfnamefont {M.}~\bibnamefont {Bal}}, \bibinfo {author} {\bibfnamefont {S.}~\bibnamefont {Garattoni}}, \bibinfo {author} {\bibfnamefont {F.}~\bibnamefont {Crisa}}, \bibinfo {author} {\bibfnamefont {D.}~\bibnamefont {Bafia}}, \bibinfo {author} {\bibfnamefont {D.}~\bibnamefont {Kurkcuoglu}}, \bibinfo {author} {\bibfnamefont {R.}~\bibnamefont {Pilipenko}}, \bibinfo {author} {\bibfnamefont {P.}~\bibnamefont {Heidler}}, \bibinfo {author} {\bibfnamefont {N.}~\bibnamefont {Bornman}}, \bibinfo {author} {\bibfnamefont {D.}~\bibnamefont {van Zanten}},  \emph {et~al.},\ }\bibfield  {title} {\enquote {\bibinfo
  {title} {Ultracoherent superconducting cavity-based multiqudit platform with error-resilient control},}\ }\href@noop {} {\bibfield  {journal} {\bibinfo  {journal} {\href{https://arxiv.org/abs/2506.03286}{arXiv:2506.03286}}\ } (\bibinfo {year} {2025})}\BibitemShut {NoStop}%
\bibitem [{\citenamefont {Harrington}\ \emph {et~al.}(2022)\citenamefont {Harrington}, \citenamefont {Mueller},\ and\ \citenamefont {Murch}}]{Harrington_NRP2022_4_660}%
  \BibitemOpen
  \bibfield  {author} {\bibinfo {author} {\bibfnamefont {P.~M.}\ \bibnamefont {Harrington}}, \bibinfo {author} {\bibfnamefont {E.~J.}\ \bibnamefont {Mueller}}, \ and\ \bibinfo {author} {\bibfnamefont {K.~W.}\ \bibnamefont {Murch}},\ }\bibfield  {title} {\enquote {\bibinfo {title} {Engineered dissipation for quantum information science},}\ }\href {\doibase 10.1038/s42254-022-00494-8} {\bibfield  {journal} {\bibinfo  {journal} {Nat. Rev. Phys.}\ }\textbf {\bibinfo {volume} {4}},\ \bibinfo {pages} {660} (\bibinfo {year} {2022})}\BibitemShut {NoStop}%
\bibitem [{\citenamefont {Udupa}\ \emph {et~al.}(2026)\citenamefont {Udupa}, \citenamefont {Hillmann}, \citenamefont {Ahmed}, \citenamefont {Smirne},\ and\ \citenamefont {Ferrini}}]{Udupa2026}%
  \BibitemOpen
  \bibfield  {author} {\bibinfo {author} {\bibfnamefont {A.}~\bibnamefont {Udupa}}, \bibinfo {author} {\bibfnamefont {T.}~\bibnamefont {Hillmann}}, \bibinfo {author} {\bibfnamefont {R.~G.}\ \bibnamefont {Ahmed}}, \bibinfo {author} {\bibfnamefont {A.}~\bibnamefont {Smirne}}, \ and\ \bibinfo {author} {\bibfnamefont {G.}~\bibnamefont {Ferrini}},\ }\bibfield  {title} {\enquote {\bibinfo {title} {Performance of rotation-symmetric bosonic codes in the presence of non-{Markovian} effects induced by random telegraph noise},}\ }\href {\doibase 10.1103/xfv4-8dq7} {\bibfield  {journal} {\bibinfo  {journal} {Phys. Rev. Research}\ }\textbf {\bibinfo {volume} {8}},\ \bibinfo {pages} {023007} (\bibinfo {year} {2026})}\BibitemShut {NoStop}%
\bibitem [{\citenamefont {Webster}\ \emph {et~al.}(2015)\citenamefont {Webster}, \citenamefont {Bartlett},\ and\ \citenamefont {Poulin}}]{Webster2015PRA_92_062309}%
  \BibitemOpen
  \bibfield  {author} {\bibinfo {author} {\bibfnamefont {P.}~\bibnamefont {Webster}}, \bibinfo {author} {\bibfnamefont {S.~D.}\ \bibnamefont {Bartlett}}, \ and\ \bibinfo {author} {\bibfnamefont {D.}~\bibnamefont {Poulin}},\ }\bibfield  {title} {\enquote {\bibinfo {title} {Reducing the overhead for quantum computation when noise is biased},}\ }\href {\doibase 10.1103/PhysRevA.92.062309} {\bibfield  {journal} {\bibinfo  {journal} {Phys. Rev. A}\ }\textbf {\bibinfo {volume} {92}},\ \bibinfo {pages} {062309} (\bibinfo {year} {2015})}\BibitemShut {NoStop}%
\bibitem [{\citenamefont {Deffner}\ and\ \citenamefont {Campbell}(2019)}]{Deffner_book_2019}%
  \BibitemOpen
  \bibfield  {author} {\bibinfo {author} {\bibfnamefont {S.}~\bibnamefont {Deffner}}\ and\ \bibinfo {author} {\bibfnamefont {S.}~\bibnamefont {Campbell}},\ }\href {\doibase 10.1088/2053-2571/ab21c6} {\emph {\bibinfo {title} {Quantum Thermodynamics: An Introduction to the Thermodynamics of Quantum Information}}}\ (\bibinfo  {publisher} {Morgan \& Claypool Publishers},\ \bibinfo {address} {San Rafael, CA},\ \bibinfo {year} {2019})\BibitemShut {NoStop}%
\end{thebibliography}

\begin{thebibliography}{54}%
\makeatletter
\providecommand \@ifxundefined [1]{%
 \@ifx{#1\undefined}
}%
\providecommand \@ifnum [1]{%
 \ifnum #1\expandafter \@firstoftwo
 \else \expandafter \@secondoftwo
 \fi
}%
\providecommand \@ifx [1]{%
 \ifx #1\expandafter \@firstoftwo
 \else \expandafter \@secondoftwo
 \fi
}%
\providecommand \natexlab [1]{#1}%
\providecommand \enquote  [1]{``#1''}%
\providecommand \bibnamefont  [1]{#1}%
\providecommand \bibfnamefont [1]{#1}%
\providecommand \citenamefont [1]{#1}%
\providecommand \href@noop [0]{\@secondoftwo}%
\providecommand \href [0]{\begingroup \@sanitize@url \@href}%
\providecommand \@href[1]{\@@startlink{#1}\@@href}%
\providecommand \@@href[1]{\endgroup#1\@@endlink}%
\providecommand \@sanitize@url [0]{\catcode `\\12\catcode `\$12\catcode `\&12\catcode `\#12\catcode `\^12\catcode `\_12\catcode `\%12\relax}%
\providecommand \@@startlink[1]{}%
\providecommand \@@endlink[0]{}%
\providecommand \url  [0]{\begingroup\@sanitize@url \@url }%
\providecommand \@url [1]{\endgroup\@href {#1}{\urlprefix }}%
\providecommand \urlprefix  [0]{URL }%
\providecommand \Eprint [0]{\href }%
\providecommand \doibase [0]{http://dx.doi.org/}%
\providecommand \selectlanguage [0]{\@gobble}%
\providecommand \bibinfo  [0]{\@secondoftwo}%
\providecommand \bibfield  [0]{\@secondoftwo}%
\providecommand \translation [1]{[#1]}%
\providecommand \BibitemOpen [0]{}%
\providecommand \bibitemStop [0]{}%
\providecommand \bibitemNoStop [0]{.\EOS\space}%
\providecommand \EOS [0]{\spacefactor3000\relax}%
\providecommand \BibitemShut  [1]{\csname bibitem#1\endcsname}%
\let\auto@bib@innerbib\@empty
\bibitem [{\citenamefont {Place}\ \emph {et~al.}(2021)\citenamefont {Place}, \citenamefont {Rodgers}, \citenamefont {Mundada}, \citenamefont {Smitham}, \citenamefont {Fitzpatrick}, \citenamefont {Leng}, \citenamefont {Premkumar}, \citenamefont {Bryon}, \citenamefont {Vrajitoarea}, \citenamefont {Sussman}, \citenamefont {Cheng}, \citenamefont {Madhavan}, \citenamefont {Babla}, \citenamefont {Le}, \citenamefont {Gang}, \citenamefont {J{\"{a}}ck}, \citenamefont {Gyenis}, \citenamefont {Yao}, \citenamefont {Cava}, \citenamefont {de~Leon},\ and\ \citenamefont {Houck}}]{Place2021}%
  \BibitemOpen
  \bibfield  {author} {\bibinfo {author} {\bibfnamefont {A.~P.~M.}\ \bibnamefont {Place}}, \bibinfo {author} {\bibfnamefont {L.~V.~H.}\ \bibnamefont {Rodgers}}, \bibinfo {author} {\bibfnamefont {P.}~\bibnamefont {Mundada}}, \bibinfo {author} {\bibfnamefont {B.~M.}\ \bibnamefont {Smitham}}, \bibinfo {author} {\bibfnamefont {M.}~\bibnamefont {Fitzpatrick}}, \bibinfo {author} {\bibfnamefont {Z.}~\bibnamefont {Leng}}, \bibinfo {author} {\bibfnamefont {A.}~\bibnamefont {Premkumar}}, \bibinfo {author} {\bibfnamefont {J.}~\bibnamefont {Bryon}}, \bibinfo {author} {\bibfnamefont {A.}~\bibnamefont {Vrajitoarea}}, \bibinfo {author} {\bibfnamefont {S.}~\bibnamefont {Sussman}}, \bibinfo {author} {\bibfnamefont {G.}~\bibnamefont {Cheng}}, \bibinfo {author} {\bibfnamefont {T.}~\bibnamefont {Madhavan}}, \bibinfo {author} {\bibfnamefont {H.~K.}\ \bibnamefont {Babla}}, \bibinfo {author} {\bibfnamefont {X.~H.}\ \bibnamefont {Le}}, \bibinfo {author} {\bibfnamefont {Y.}~\bibnamefont {Gang}}, \bibinfo {author} {\bibfnamefont
  {B.}~\bibnamefont {J{\"{a}}ck}}, \bibinfo {author} {\bibfnamefont {A.}~\bibnamefont {Gyenis}}, \bibinfo {author} {\bibfnamefont {N.}~\bibnamefont {Yao}}, \bibinfo {author} {\bibfnamefont {R.~J.}\ \bibnamefont {Cava}}, \bibinfo {author} {\bibfnamefont {N.~P.}\ \bibnamefont {de~Leon}}, \ and\ \bibinfo {author} {\bibfnamefont {A.~A.}\ \bibnamefont {Houck}},\ }\bibfield  {title} {\enquote {\bibinfo {title} {New material platform for superconducting transmon qubits with coherence times exceeding 0.3 milliseconds},}\ }\href {\doibase 10.1038/s41467-021-22030-5} {\bibfield  {journal} {\bibinfo  {journal} {Nat. Commun.}\ }\textbf {\bibinfo {volume} {12}},\ \bibinfo {pages} {1779} (\bibinfo {year} {2021})}\BibitemShut {NoStop}%
\bibitem [{\citenamefont {Gardiner}\ and\ \citenamefont {Collett}(1985)}]{Gardiner1985}%
  \BibitemOpen
  \bibfield  {author} {\bibinfo {author} {\bibfnamefont {C.~W.}\ \bibnamefont {Gardiner}}\ and\ \bibinfo {author} {\bibfnamefont {M.~J.}\ \bibnamefont {Collett}},\ }\bibfield  {title} {\enquote {\bibinfo {title} {Input and output in damped quantum systems: Quantum stochastic differential equations and the master equation},}\ }\href {\doibase 10.1103/PhysRevA.31.3761} {\bibfield  {journal} {\bibinfo  {journal} {Phys. Rev. A}\ }\textbf {\bibinfo {volume} {31}},\ \bibinfo {pages} {3761} (\bibinfo {year} {1985})}\BibitemShut {NoStop}%
\bibitem [{\citenamefont {Shinozuka}\ and\ \citenamefont {Jan}(1972)}]{shinozuka1972}%
  \BibitemOpen
  \bibfield  {author} {\bibinfo {author} {\bibfnamefont {M.}~\bibnamefont {Shinozuka}}\ and\ \bibinfo {author} {\bibfnamefont {C.-M.}\ \bibnamefont {Jan}},\ }\bibfield  {title} {\enquote {\bibinfo {title} {Digital simulation of random processes and its applications},}\ }\href {\doibase 10.1016/0022-460X(72)90600-1} {\bibfield  {journal} {\bibinfo  {journal} {J. Sound Vib.}\ }\textbf {\bibinfo {volume} {25}},\ \bibinfo {pages} {111} (\bibinfo {year} {1972})}\BibitemShut {NoStop}%
\bibitem [{\citenamefont {Khaneja}\ \emph {et~al.}(2005)\citenamefont {Khaneja}, \citenamefont {Reiss}, \citenamefont {Kehlet}, \citenamefont {Schulte-Herbr{\"{u}}ggen},\ and\ \citenamefont {Glaser}}]{Khaneja2005}%
  \BibitemOpen
  \bibfield  {author} {\bibinfo {author} {\bibfnamefont {N.}~\bibnamefont {Khaneja}}, \bibinfo {author} {\bibfnamefont {T.}~\bibnamefont {Reiss}}, \bibinfo {author} {\bibfnamefont {C.}~\bibnamefont {Kehlet}}, \bibinfo {author} {\bibfnamefont {T.}~\bibnamefont {Schulte-Herbr{\"{u}}ggen}}, \ and\ \bibinfo {author} {\bibfnamefont {S.~J.}\ \bibnamefont {Glaser}},\ }\bibfield  {title} {\enquote {\bibinfo {title} {Optimal control of coupled spin dynamics: design of {NMR} pulse sequences by gradient ascent algorithms},}\ }\href {\doibase 10.1016/j.jmr.2004.11.004} {\bibfield  {journal} {\bibinfo  {journal} {J. Magn. Reson.}\ }\textbf {\bibinfo {volume} {172}},\ \bibinfo {pages} {296} (\bibinfo {year} {2005})}\BibitemShut {NoStop}%
\bibitem [{\citenamefont {Deng}\ \emph {et~al.}(2024)\citenamefont {Deng}, \citenamefont {Li}, \citenamefont {Chen}, \citenamefont {Ni}, \citenamefont {Cai}, \citenamefont {Mai}, \citenamefont {Zhang}, \citenamefont {Zheng}, \citenamefont {Yu}, \citenamefont {Zou}, \citenamefont {Liu}, \citenamefont {Yan}, \citenamefont {Xu},\ and\ \citenamefont {Yu}}]{deng2024}%
  \BibitemOpen
  \bibfield  {author} {\bibinfo {author} {\bibfnamefont {X.}~\bibnamefont {Deng}}, \bibinfo {author} {\bibfnamefont {S.}~\bibnamefont {Li}}, \bibinfo {author} {\bibfnamefont {Z.-J.}\ \bibnamefont {Chen}}, \bibinfo {author} {\bibfnamefont {Z.}~\bibnamefont {Ni}}, \bibinfo {author} {\bibfnamefont {Y.}~\bibnamefont {Cai}}, \bibinfo {author} {\bibfnamefont {J.}~\bibnamefont {Mai}}, \bibinfo {author} {\bibfnamefont {L.}~\bibnamefont {Zhang}}, \bibinfo {author} {\bibfnamefont {P.}~\bibnamefont {Zheng}}, \bibinfo {author} {\bibfnamefont {H.}~\bibnamefont {Yu}}, \bibinfo {author} {\bibfnamefont {C.-L.}\ \bibnamefont {Zou}}, \bibinfo {author} {\bibfnamefont {S.}~\bibnamefont {Liu}}, \bibinfo {author} {\bibfnamefont {F.}~\bibnamefont {Yan}}, \bibinfo {author} {\bibfnamefont {Y.}~\bibnamefont {Xu}}, \ and\ \bibinfo {author} {\bibfnamefont {D.}~\bibnamefont {Yu}},\ }\bibfield  {title} {\enquote {\bibinfo {title} {Quantum-enhanced metrology with large {Fock} states},}\ }\href {\doibase 10.1038/s41567-024-02619-5}
  {\bibfield  {journal} {\bibinfo  {journal} {Nat. Phys.}\ }\textbf {\bibinfo {volume} {20}},\ \bibinfo {pages} {1874} (\bibinfo {year} {2024})}\BibitemShut {NoStop}%
\end{thebibliography}

%

\end{document}